\documentclass[use AMS,use natbib]{mn2e}

\usepackage{amsmath}
\usepackage{amssymb}
\usepackage[T1]{fontenc}
\usepackage[latin1]{inputenc}
\usepackage{framed} 
\usepackage{color}
\usepackage{multirow}
\usepackage{textcomp}
\usepackage{units}
\usepackage{units}
\usepackage{tabularx}
\usepackage{array}
\usepackage{color}
\usepackage{textcomp}
\usepackage[]{epic,eepic,latexsym,epsf}
\usepackage{graphicx}

\begin{document}

%%%%%%%%%%%%%%%%%%%%%%%%%%%%%%%%%%%%%%%%%%%%%%%%%%%%%%%%%%%%%%%%%%%%%%%%%%%%%%%%%%%%%%%%%%%%%%%%%%%%%%%%%%%%%%%%%%%%%%%%%%
%%%%%%%%%%%%%%%%%%%%%%%% Title / Abstract %%%%%%%%%%%%%%%%%%%%%%%%%%%%%%%%%%%%%%%%%%%%%%%%%%%%%%%%%%%%%%%%%%%%%%%%%%%%%%%%
%%%%%%%%%%%%%%%%%%%%%%%%%%%%%%%%%%%%%%%%%%%%%%%%%%%%%%%%%%%%%%%%%%%%%%%%%%%%%%%%%%%%%%%%%%%%%%%%%%%%%%%%%%%%%%%%%%%%%%%%%%

\title{Synthetic X-ray and radio maps for two different models of Stephan's Quintet}

\author[A. Geng et al.]{Annette Geng$^{1}$\thanks{E-mail: annette.geng@uni-konstanz.de}, Alexander M. Beck$^{2,3}$, Klaus Dolag$^{2,4}$, Florian B\"{u}rzle$^{1}$, \newauthor Marcus C. Beck$^{1}$, Hanna Kotarba$^{2}$, and Peter Nielaba$^{1}$ \newauthor\\
$^{1}$University of Konstanz, Department of Physics, Universit\"{a}tsstr. 10, 78457 Konstanz, Germany\\
$^{2}$University Observatory Munich, Scheinerstr. 1, 81679 Munich, Germany\\
$^{3}$Max Planck Institute for Extraterrestrial Physics, Giessenbachstrasse, 85748 Garching, Germany\\ 
$^{4}$Max Planck Institute for Astrophysics, Karl-Schwarzschild-Str. 1, 85741 Garching, Germany}

\maketitle

\begin{abstract}
We present simulations of the compact galaxy group Stephan's Quintet (SQ) including magnetic fields, performed with the N-body/smoothed particle hydrodynamics (SPH) code \textsc{Gadget}. The simulations include radiative cooling, star formation and supernova feedback. Magnetohydrodynamics (MHD) is implemented using the standard smoothed particle magnetohydrodynamics (SPMHD) method. We adapt two different initial models for SQ based on Renaud et al. and Hwang et al., both including four galaxies (NGC 7319, NGC 7320c, NGC 7318a and NGC 7318b). Additionally, the galaxies are embedded in a magnetized, low density intergalactic medium (IGM). The ambient IGM has an initial magnetic field of $10^{-9}$ G and the four progenitor discs have initial magnetic fields of $10^{-9} - 10^{-7}$ G. We investigate the morphology, regions of star formation, temperature, X-ray emission, magnetic field structure and radio emission within the two different SQ models. In general, the enhancement and propagation of the studied gaseous properties (temperature, X-ray emission, magnetic field strength and synchrotron intensity) is more efficient for the SQ model based on Renaud et al., whose galaxies are more massive, whereas the less massive SQ model based on Hwang et al. shows generally similar effects but with smaller efficiency. We show that the large shock found in observations of SQ is most likely the result of a collision of the galaxy NGC 7318b with the IGM. This large group-wide shock is clearly visible in the X-ray emission and synchrotron intensity within the simulations of both SQ models. The order of magnitude of the observed synchrotron emission within the shock front is slightly better reproduced by the SQ model based on Renaud et al., whereas the distribution and structure of the synchrotron emission is better reproduced by the SQ model based on Hwang et al..
\end{abstract}


%%%%%%%%%%%%%%%%%%%%%%%%%%%%%%%%%%%%%%%%%%%%%%%%%%%%%%%%%%%%%%%%%%%%%%%%%%%%%%%%%%%%%%%%%%%%%%%%%%%%%%%%%%%%%%%%%%%%%%%%%%
%%%%%%%%%%%%%%%%%%%%%%%% Introduction %%%%%%%%%%%%%%%%%%%%%%%%%%%%%%%%%%%%%%%%%%%%%%%%%%%%%%%%%%%%%%%%%%%%%%%%%%%%%%%%%%%%
%%%%%%%%%%%%%%%%%%%%%%%%%%%%%%%%%%%%%%%%%%%%%%%%%%%%%%%%%%%%%%%%%%%%%%%%%%%%%%%%%%%%%%%%%%%%%%%%%%%%%%%%%%%%%%%%%%%%%%%%%%

\section[]{Introduction}

\begin{figure}
\begin{center}
\includegraphics[width=0.9\columnwidth]{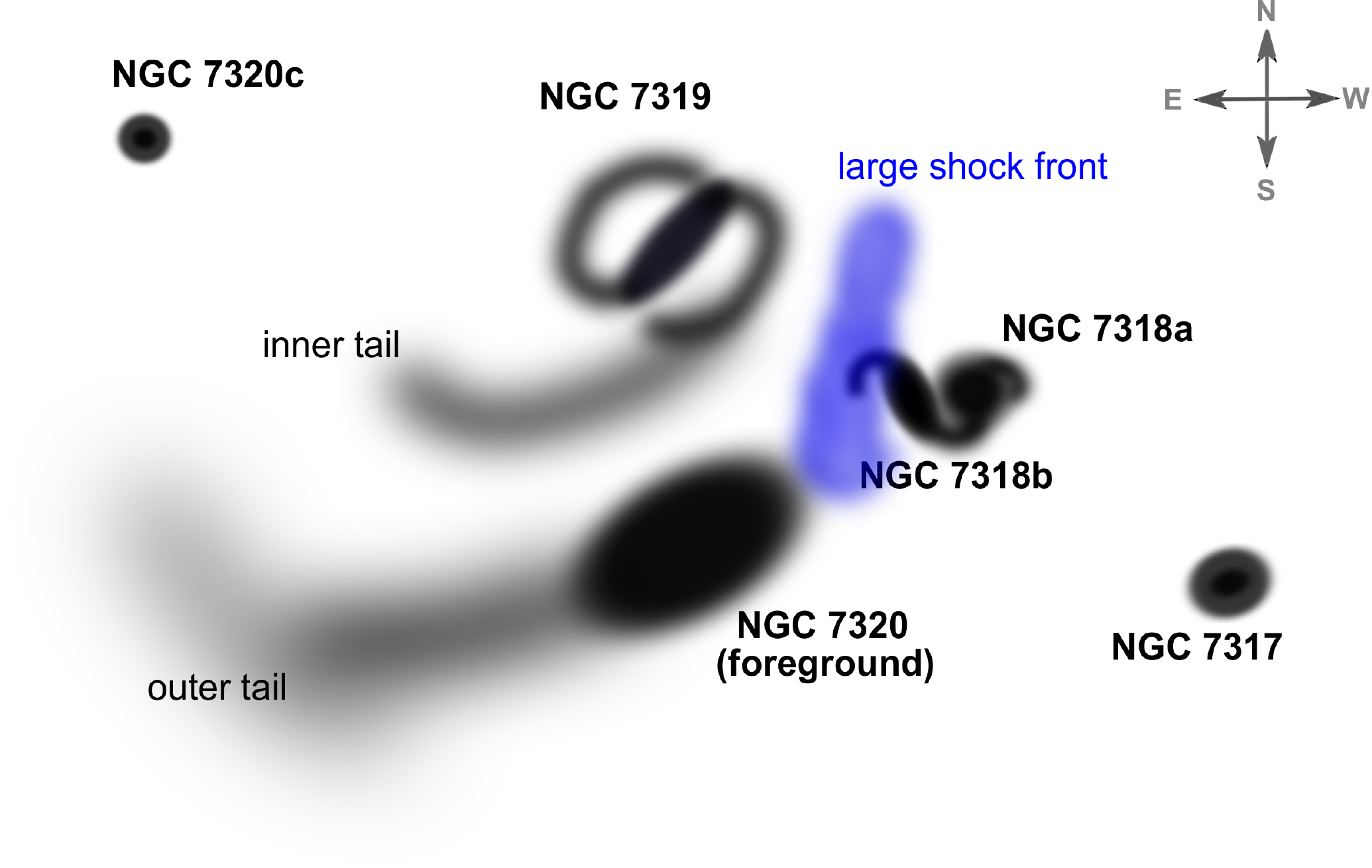}
\end{center}
\caption{\small{Schematic view of the main morphological features of SQ. Historically, SQ consists of the galaxies NGC 7319, NGC 7318a/b, NGC 7320 and NGC 7317. Today, however, only the galaxies NGC 7319, NGC 7318a/b, NGC 7317 and additionally NGC 7320c are observed to form an interdependent galaxy group. As indicated in the upper right (grey), north is up and east is to the left.}}
\label{morph}
\end{figure}

Stephan's Quintet (\citealt{St77}; hereafter SQ), also known as Hickson Compact Group 92 \citep{Hi82}, is the first discovered compact galaxy group. It is located in the constellation Pegasus. SQ consists of five galaxies (NGC 7319, NGC 7318a, NGC 7318b, NGC 7317, and NGC 7320, cf. also Fig. \ref{morph}) with an estimated distance to earth of $\approx$ 94 Mpc \citep{MoMa98,ApXu06}. SQ is famous for a physical adjacency between four of the galaxies (NGC 7319, 7318a, 7318b, and 7317), whereby strong interactions between three of these members are apparently causing tidal tails, a strong group-wide shock visible in X-ray \citep{PiTr97,SuRo01} and radio emission \citep{AlHa72,XuLu03}, and a region of active star formation northern of the colliding galaxies \citep{Xu05}. The fifth galaxy, NGC 7320, is observed to be a much closer foreground galaxy and is not part of the interacting group \citep{Sh74,AlSu80,MoSu97}. It has a recessional velocity of $\approx$ 740 km s$^{-1}$ \citep{FaKu99}. Three of the four physically related galaxies, the main galaxy NGC 7319 as well as NGC 7317 and NGC 7318a, have a similar recessional velocity  of $\approx$ 6640-6670 km s$^{-1}$ \citep{FeGa11} and represent the core of the compact group (c.f. Fig. \ref{morph}). NGC 7318b is observed to be a high-speed intruder ($\approx$ 5770 km s$^{-1}$) and seems to interact with the IGM within the main system for the first time \citep{MoSu97,FeGa11}. Finally, there is a sixth galaxy, NGC 7320c, which shows a similar recessional velocity ($\approx$ 5990 km s$^{-1}$) as the core of the compact group \citep{FeGa11} and is therefore suggested to interact with NGC 7319 \citep{MoSu97}, because it reveals connected tidal features in the eastern large tidal tail (cf. the outer tail in Fig. \ref{morph}). Therefore, it is most likely also part of the compact group \citep{Ar73}.

Radio observations reveal the presence of magnetic fields in most late-type galaxies of the local Universe. The field strengths range from a few $\mu$G in isolated quiet galaxies up to about 50 $\mu$G in starburst galaxies (see e.g. \citealt{Hu86,KlWi88,FiAl93,BeBr96,ChKn03}). The intergalactic magnetic field is usually estimated to be less than 0.01 $\mu$G \citep[e.g.][]{KrBe08}. Galaxy interactions can cause magnetic fields which are much stronger compared to individual galaxies (\citealt{Be05,DrCh11}). In this context, the SQ is particularly interesting because it shows both galaxy-galaxy and galaxy-IGM interactions. The latter is mainly visible on the basis of the prominent ridge of X-ray and radio emission crossing the system. \cite{XuLu03} estimate the minimum-energy magnetic field strength within the shock region of SQ to $\approx$ 10 $\mu$G.

The morphology of an interacting galaxy system strongly depends on the initial properties of the progenitors. Numerical simulations can provide insights into the properties of the initial progenitor galaxies, as well as possible formation scenarios. So far, simulations of galaxy interactions were predominantly focusing on stellar dynamics, gas flows, star formation, supermassive black holes and feedback from stars and black holes \citep[e.g.][]{DiSp05,SpDi05a,SpDi05b,RoBu06,CoJo08,JoNa09}. However, over the past few years, the interest in simulations of magnetic fields in galactic environments has grown substantially \citep[e.g.][]{KoLe09,KoLe11,DuTe10,GeKo12}.
Synthetic radio maps of the interacting system of the Antennae galaxies were recently presented by \citet{KoKa10}. They showed radio and polarization maps calculated from their SPMHD simulations to be in good morphological and quantitative agreement with observations, implying that the simulations have the capability to follow the magnetic field evolution in a highly interacting nonlinear environment.

Despite an enormous number of observational studies of SQ, revealing different features at different wavelengths, numerical simulations of this system are very rare, due to the difficulties of many free parameters in modeling such a complex interacting system. The simulated models for SQ commonly exclude the foreground galaxy NGC 7320 but include NGC 7320c as a member of the galaxy group. Furthermore, NGC 7317 does not show any visible features of an ongoing or past interaction. Therefore, it is usually not taken into account in the simulations. As a first basis for more detailed studies, \citet{ReAp10} performed collisionless gravitational N-body simulations proposing a possible formation scenario for SQ. Their simulation is mainly focusing on reproducing the stellar large-scale structure of SQ. Recently, \citet{HwSt12} presented an extended model including a gaseous component and different galaxy models, suggesting a different formation history. The global morphology of the system is well represented in their simulation, supporting also the idea that the large-scale shock within the system is caused by the interaction of NGC 7318b with the intragroup IGM.

However, until now, detailed numerical studies of the SQ concerning star formation, temperature, magnetic fields, X-ray and radio emission are still missing. Therefore, within the presented work, we perform SPMHD studies of SQ on the basis of the two existing models by \citet{ReAp10} and \citet{HwSt12}. Thereby, we place a particular focus on the properties of the gaseous component, i.e. on the star formation rate, the temperature, and the magnetic field. We also investigate the presence of the shock front in the X-ray and radio emission. Our synthetic radio maps are calculated for different frequencies, allowing a better comparability with observations. 

The paper is organized as follows. In section 2 we introduce the main observational properties of SQ. Section 3 briefly summerizes the existing models of SQ and presents our extended initial conditions. The numerical method is described in section 4. In section 5 we present the results of our simulated SQ models, including star formation regions, temperature distribution, X-ray emission, magnetic field structure, radio emission and polarization maps. We discuss the two different SQ models in section 6. Finally, we summerize and discuss our findings in section 7.

%%%%%%%%%%%%%%%%%%%%%%%%%%%%%%%%%%%%%%%%%%%%%%%%%%%%%%%%%%%%%%%%%%%%%%%%%%%%%%%%%%%%%%%%%%%%%%%%%%%%%%%%%%%%%%%%%%%%%%%%%%
%%%%%%%%%%%%%%%%%%%%%%%% Observations %%%%%%%%%%%%%%%%%%%%%%%%%%%%%%%%%%%%%%%%%%%%%%%%%%%%%%%%%%%%%%%%%%%%%%%%%%%%%%%%%%%%
%%%%%%%%%%%%%%%%%%%%%%%%%%%%%%%%%%%%%%%%%%%%%%%%%%%%%%%%%%%%%%%%%%%%%%%%%%%%%%%%%%%%%%%%%%%%%%%%%%%%%%%%%%%%%%%%%%%%%%%%%%

\section[]{Observational properties of the interacting group SQ}

SQ is probably the most studied compact galaxy group. The system shows interesting tidal features resulting from one or more strong interactions of the galaxies, namely the outer tail and the inner tail (cf. Fig. \ref{morph}). It was suggested that the outer tail could have formed earlier in time by an interaction event between NGC 7319 and an intruder, whereas the development of the inner tail seems to be caused by a more recent event \citep[e.g.][]{MoSu97,Xu05}, therefore the tails are sometimes also referred to the old and young tails. Note, however, that there are three different formation scenarios concerning the tails, which are still under discussion (see end of this section). The outer tail is extending to the southeast of NGC 7319 and is mostly covered by the foreground galaxy NGC 7320. The inner tail runs parallel to the outer tail below NGC 7319 and shows a higher surface brightness \citep{FeGa11} and an active star formation \citep{Xu05}. At its eastern end, the inner tail is pointing towards NGC 7320c. North of NGC 7318a and NGC 7318b there are two tidal arms extending from the galaxies, which belong to a region of active star formation \citep{GaCh01,Xu05}.
A region of enhanced X-ray and radio emission is observed between NGC 7319 and NGC 7318b. This region is widely interpreted as a large shock front (cf. Fig. \ref{morph}) resulting from a collision of the high-speed intruder NGC 7318b with the cold IGM \citep{ShAl84}.

\begin{figure}
\includegraphics[width=1.\columnwidth]{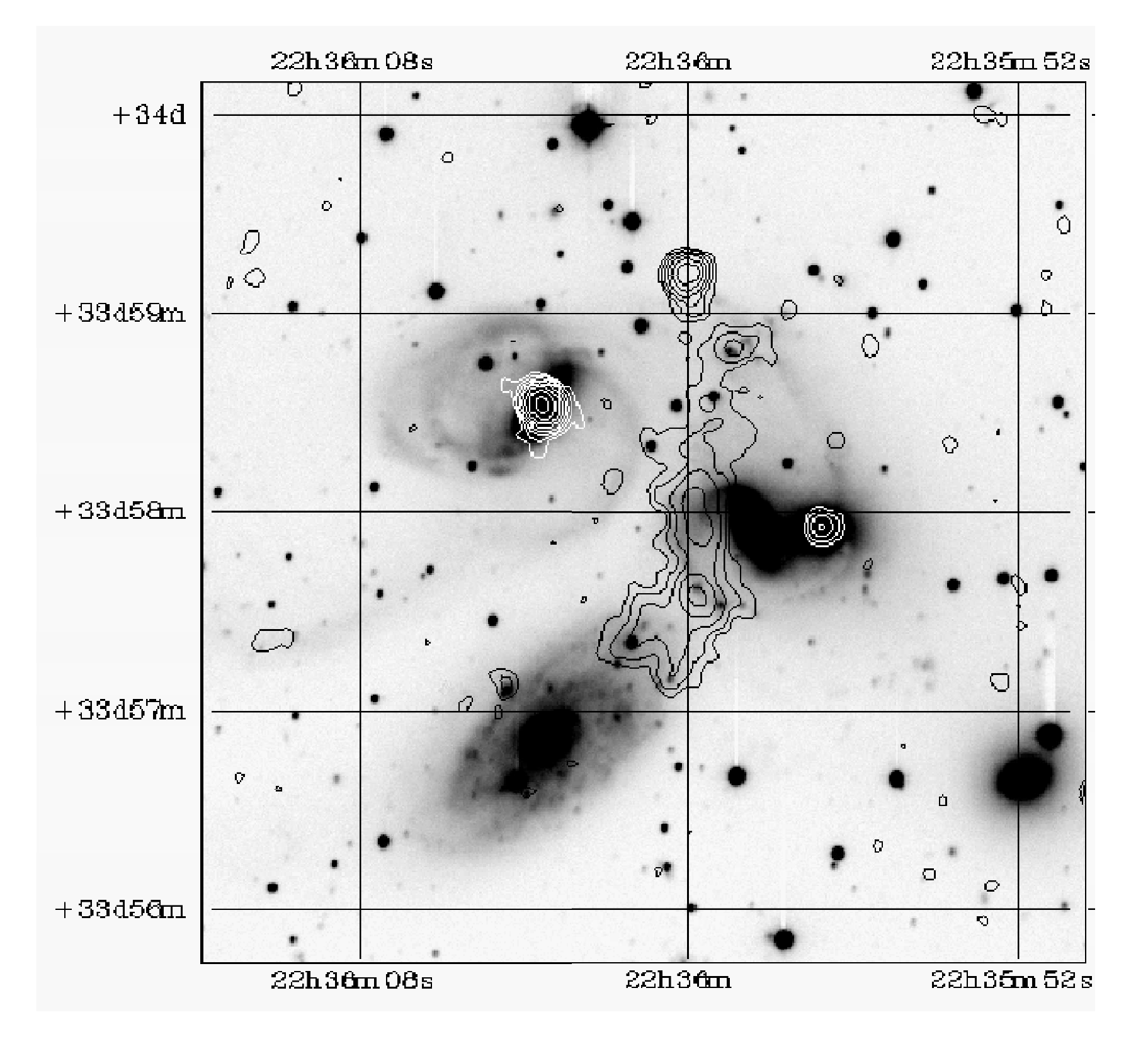}
\caption{\small{R-band CCD image overlaid with contours of the radio continuum at 4.86 GHz (total radio emission) observed with the VLA. The lowest contour is 50 $\mu$Jy beam$^{-1}$ with a further spacing of 2 in ratio. Plot from \citet{XuLu03} reproduced by the friendly permission of the author and the AAS.}}
\label{Xu}
\end{figure}

Originally, the large shock was discovered in the 21 cm Westerborg radio continuum map by \citet{AlHa72} (see Fig. \ref{Xu}). Later, the prominent ridge was shown to be also present on observed X-ray maps \citep{PiTr97,SuRo01}. Due to the additional, albeit faint, evidence of the ridge in observations within the optical wavelength, \citet{XuLu03} suggested that the shock front contains ionized gas and cold dust as well as hot thermal electrons, relativistic electrons and strong magnetic fields. 
However, the shock region between NGC 7319 and NGC 7318b is not the only radio source in SQ (cf. Fig. \ref{Xu}): further observed regions of enhanced radio emission are the innermost parts of NGC 7319 and NGC 7318a. Also, the region north of NGC 7318a and NGC 7318b, which is observed to have a high star formation rate (see below), is also found to emit synchrotron radiation \citep{XuLu03}. 

The star formation activity in SQ is believed to be triggered by the interactions and can be measured in UV \citep{Xu05}. The system reveals several regions of active star formation ($\ge 1$ M$_{\sun}$ yr$^{-1}$), such as the inner disc of NGC 7318b, the disc of NGC 7319 and the star-forming region in the north of the galaxy pair NGC 7318a/b. \citet{Cl10} found active star formation regions in the shock to be located in the intruder galaxy and at either end of the radio shock, whereas the shock ridge itself seems to host only regions of weak star formation. Additional regions of weak star formation ($\le 1$ M$_{\sun}$ yr$^{-1}$) are found in the outer disc of NGC 7318b, within the inner and outer tails and in the overlap region of the tail \citep{Xu05}.

The outer and the inner tails are the result of interactions within SQ. Therefore, any model suggesting an interaction history for SQ should be able to reproduce the structure and age of these features. Three different possible interaction scenarios discussed in the literature meet this requirement. 
The first scenario suggests a formation of the outer tail at an earlier time than the formation of the inner tail. \citet{Xu05} proposed that the outer tail is the result of an encounter between the galaxies NGC 7320c and NGC 7319. The inner tail is suggested to have formed later due to an encounter of NGC 7319 with NGC 7318a. 
The second possible scenario also suggests a formation of the tails at different times but both resulting from passages of NGC 7320c over NGC 7319. Based on the fact that both the outer and the inner tails are directed towards NGC 7320c, \citet{MoMa98} and also \citet{SuRo01} proposed a bound orbit of NGC 7320c around NGC 7319, where the first interaction created the outer and a subsequent passage the inner tail.
The third scenario assumes a formation of both tails by only one interaction event of NGC 7320c and NGC 7319 \citep[e.g.][]{HwSt12} thus suggesting the same formation age for the two tails. Although the common opinion favors different formation times of the tails due to i.e. different star formation rates and cluster ages within the tails (e.g. \citealt{FeGa11}), this issue is still under discussion \citep{HwSt12}.

%%%%%%%%%%%%%%%%%%%%%%%%%%%%%%%%%%%%%%%%%%%%%%%%%%%%%%%%%%%%%%%%%%%%%%%%%%%%%%%%%%%%%%%%%%%%%%%%%%%%%%%%%%%%%%%%%%%%%%%%%%
%%%%%%%%%%%%%%%%%%%%%%%% Initial Conditions %%%%%%%%%%%%%%%%%%%%%%%%%%%%%%%%%%%%%%%%%%%%%%%%%%%%%%%%%%%%%%%%%%%%%%%%%%%%%%
%%%%%%%%%%%%%%%%%%%%%%%%%%%%%%%%%%%%%%%%%%%%%%%%%%%%%%%%%%%%%%%%%%%%%%%%%%%%%%%%%%%%%%%%%%%%%%%%%%%%%%%%%%%%%%%%%%%%%%%%%%

\section[]{Initial Conditions}

\begin{table*}
\caption{Parameters of initial galaxy setup for both models of SQ}
\begin{center}
\renewcommand{\arraystretch}{1.2}
\begin{tabular}{lllll}
\hline\hline  
\multicolumn{5}{c}{\textsc{Galaxy Parameters}} \\ \hline\hline
\hspace{12mm}&NGC7319\hspace{5mm}&NGC7320c\hspace{5mm}&NGC7318a\hspace{5mm}&NGC7318b\hspace{5mm}\\\hline
Concentration$^{a}$ CC&12&20/8&12&12\\
Spin parameter $\lambda$&0.1&0.15&0.1&0.1\\
Disk mass fraction$^{b}$ $m_{\text{d}}$&0.125 $\cdot$ M$_{\text{tot}}$&0.05 $\cdot$ M$_{\text{tot}}$&0.125 $\cdot$ M$_{\text{tot}}$&0.125 $\cdot$ M$_{\text{tot}}$\\ 
Bulge mass fraction$^{b}$ $m_{\text{b}}$&0.0625 $\cdot$ M$_{\text{tot}}$&0.0214 $\cdot$ M$_{\text{tot}}$&0.0625 $\cdot$ M$_{\text{tot}}$&0.0625 $\cdot$ M$_{\text{tot}}$\\
Disk spin fraction $j_{\text{d}}$&0.125&0.05&0.125&0.125\\
Gas fraction f&0.2&0.2&0.2&0.2\\
Disk height$^{b}$ $z_{0}$&0.2 $\cdot$ $l_{\text{d}}$&0.2 $\cdot$ $l_{\text{d}}$&0.2 $\cdot$ $l_{\text{d}}$&0.2 $\cdot$ $l_{\text{d}}$\\
Bulge size$^{b}$ $l_{\text{b}}$&0.2 $\cdot$ $l_{\text{d}}$&0.2 $\cdot$ $l_{\text{d}}$&0.2 $\cdot$ $l_{\text{d}}$&0.2 $\cdot$ $l_{\text{d}}$\\
Scale length of extended gas disc$^{b}$ $l_{\text{g}}$&6 $\cdot$ $l_{\text{d}}$&6 $\cdot$ $l_{\text{d}}$&6 $\cdot$ $l_{\text{d}}$&6 $\cdot$ $l_{\text{d}}$\\\\
\multicolumn{5}{c}{\textsc{SQ model A}} \\
Initial coordinates (x,y,z) [kpc/h]&(0.0, 0.0, 0.0)&(19.92, 10.45, 19.92)&(-83.12, 0.0, 38.70)&(-14.04, 17.64, -217.19)\\
Initial velocities ($v_{x}$,$v_{y}$,$v_{z}$) [km/s]&(0.0, 0.0, 0.0)&(-620.0, 232.5, -387.5)&(465.0, -46.5, -93.0)&(218.0, 0.0, 1025.0)\\
Disc orientation ($\theta$, $\phi$)&(0, 0)&(0, 0)&(180, 23)&(0, -23)\\\\
\multicolumn{5}{c}{\textsc{SQ model B}} \\
Initial coordinates (x,y,z) [kpc/h]&(0.0, 0.0, 0.0)&(8.9, -10.9, 10.9)&(-50.0, 7.1, -40.0)&(15.0, 1.4, -241.4)\\
Initial velocities ($v_{x}$,$v_{y}$,$v_{z}$) [km/s]&(0.0, 0.0, 0.0)&(35.9, 79.5, -77.5)&(100.0, -27.0, -92.5)&(20.0, -7.5, 350.0)\\
Disc orientation ($\theta$, $\phi$)&(0, 0)&(0, 0)&(180, 0)&(180, 0)\\\\
\hline\hline

\multicolumn{5}{l}{\scriptsize{(a) Second column: model A / model B.}}\\
\multicolumn{5}{l}{\scriptsize{(b) M$_{\text{tot}}$ and l$_{\text{d}}$ are given in Table 2.}}

\end{tabular}
\end{center}
\label{TG}
\end{table*}
The first attempt towards a morphologically adequate representation of SQ in simulations was made by \citet{ReAp10}. They performed a large number of collisionless N-body simulations to find initial parameters for the four progenitor galaxies, including initial positions and velocities. Starting with these initial parameters, the system  undergoes a number of interactions resulting in a morphological structure comparable to observational findings. The large-scale configuration of the tidal features and the galaxies is generally well represented. However, these simulations are purely gravitational and therefore not suitable for more detailed studies of e.g. intergalactic gas properties, shocks, star formation activity, magnetic fields, etc. Hence, \citet{ReAp10} suggest to use their models as a basis for more complex simulations of SQ.

Recently, \citet{HwSt12} presented a further model, including also a gaseous component. They performed restricted three-body/SPH simulations of the SQ system using different models for the progenitor galaxies and a different formation scenario compared to \citet{ReAp10}. Yet, they were also able to reproduce the main tidal features of the system. Furthermore, they found indications supporting the hypothesis that the large shock between NGC 7319 and NGC 7318b has been caused by a high-speed collision of NGC 7318b and the IGM. They also studied the behaviour of gas clouds within the shocked region developing after the collision. They found a continuing production of small shocks in this region over a time span of $\approx 10^{7}$ yrs. However, they were not able to study the star formation history and the gas temperature in more detail, because their treatment of heating and cooling was not accurate enough to draw reliable conclusions. Moreover, they did not perform any investigations concerning X-ray emission, magnetic fields and radio emission.

\begin{table*}
 \begin{minipage}{126mm}
  \caption{Galaxy Model Parameters.}
\begin{center}
\renewcommand{\arraystretch}{1.2}
\begin{tabular}{lrrrrrrr}
\hline\hline  
Model&M$_{\text{tot}}$&R$_{200}$&\hspace{0.15cm}$l_{\text{d}}$&$N_{\text{halo}}^{\text{a}}$ & $N_{\text{disc}}^{\text{b}}$ & $N_{\text{gas}}^{\text{c}}$ & $N_{\text{bulge}}^{\text{d}}$\\
& \begin{footnotesize}[$10^{10}M_{\sun}$] \end{footnotesize}&\begin{footnotesize}[kpc/$h$]\end{footnotesize} &\begin{footnotesize}[kpc/$h$]\end{footnotesize} &&&&\\\hline\\
\multicolumn{7}{c}{\textsc{SQ model A}} \\
NGC7319&258.1&199.0&7.1&706 900&869 600&217 400&544 000\\
NGC7320c&44.8&111.0&5.9&141 380&60 952&15 238&32 610\\
NGC7318a&88.0&139.0&5.0&240 346&295 660&73 920&184 960\\
NGC7318b&46.0&112.0&4.0&127 242&156 530&39 130&97 920\\\\
\multicolumn{7}{c}{\textsc{SQ model B}} \\
NGC7319&12.5&63.0&2.6&737 456&907 636&226 908&567 272\\
NGC7320c&2.5&42.5&3.6&169 780&73 136&18 284&39 128\\
NGC7318a&8.2&63.0&2.3&483 364&594 908&148 728&371 820\\
NGC7318b&7.1&60.0&2.2&418 008&514 472&128 620&321 544\\\\\hline\hline
\multicolumn{8}{l}{\scriptsize{(a) collisionless particles within dark matter halo. \hspace{0.3cm} (b) collisionless particles within disc.}}\\ \multicolumn{8}{l}{\scriptsize{(c) gas particles within disc. \hspace{2.75cm}  (d) collisionless particles within bulge.}}\\
\end{tabular}
\end{center}
\end{minipage}
\label{TN}
\end{table*}

Both of the models include the four strongly interacting galaxies NGC 7319, NGC 7320c, NGC 7318a and NGC 7318b. The galaxies NGC 7317 and NGC 7320 were not considered, as the influence of NGC 7317 on the other group members is negligible at present time \citep{ReAp10} and NGC 7320 is observed to be an unrelated foreground galaxy \citep{Sh74,AlSu80,MoSu97}.

Both \citet{ReAp10} and \citet{HwSt12} have considered the three different formation scenarios described in section 2. \citet{ReAp10} found the first scenario to best represent the present-day morphology of SQ. They also found that the second scenario does not work for the simulations at all. In contrast, \citet{HwSt12} found best results using the third scenario. This is mainly because for the first scenario they were not able to reconstruct the tails one after another and to adjust the orbit of NGC 7318a in such a way that a good representation of the observations is reached. They ascribe the differences of the results mainly to the more extended halo potentials and also to the differences between the purely gravitational N-body simulations and those including gas physics using SPH \citep{HwSt12}.

Therefore, we chose in each case the best model, i.e. the first scenario model of \citet{ReAp10} and the third scenario model of \citet{HwSt12}, and used them as a basis for SPMHD studies of SQ. The sequence of interactions within the model of \citet{ReAp10} is the following: NGC 7320c undergoes a collision with NGC 7319, then 7318a interacts with the already disturbed galaxy NGC 7319 and finally the high-speed intruder NGC 7318b hits the system. For the model of \citet{HwSt12}, the interaction history is different: NGC 7320c performs a close passage around NGC 7319, then the galaxies NGC 7318a and NGC 7318b undergo a collision behind the orbital plane of the main system followed by a collision of the high-speed intruder NGC 7318b with the IGM material west of NGC 7319.

Modifications of the galaxy models and the initial positions and velocities were required due to the additional inclusion of a gaseous component and an ambient IGM. We note that we did not intend to reproduce the models of \citet{ReAp10} and \citet{HwSt12} in detail but rather to use them as basis for our studies of magnetic fields, X-ray and radio emission in SQ. In the following sections we provide details on our galaxy models and initial positions and velocities.

\subsection{Galaxy models}

\citet{ReAp10} set up their galaxies to be composed of an exponential disc, a bulge and a dark matter halo (all consisting of collisionless gravitational N-body particles) using a method based on \citet{He93}. As NGC 7320c is assumed to be spherically symmetric it is only made up of a halo and a bulge component.

For our first SQ model, we adopt the total masses as well as the percental masses of bulge, halo and disc (if present). The galaxies are set up using the method described by \citet{SpDi05b}, which is also based on the Hernquist method \citep{He93}. This method allows for a galaxy model consisting of a cold dark matter halo, an exponential stellar disc, a stellar bulge (all of these components being collisionless gravitational N-body particles) and an exponential gaseous disc (SPH particles). We included a gaseous disc component using a disc gas fraction of f = 0.2 to all of the galaxy models. The number of particles and thus the resolution were highly increased. The parameters of our galaxy models are given in Table \ref{TG}, whereby masses are given in units of the total galactic mass M$_{\text{tot}}$ and scale lengths in units of the stellar disc scale length $l_{\text{d}}$ (M$_{\text{tot}}$ and $l_{\text{d}}$ are given in Table 2). 

The galaxy models of \citet{HwSt12} are composed of a dark matter halo and a disc containing star as well as gas particles. NGC 7318a was indeed set up with a disc, but the angular and random velocities are representing an elliptical. Surprisingly, the total masses of the galaxies are roughly a factor of 10 smaller compared to the SQ model of \cite{ReAp10}. Also, the mass ratios of the galaxies among each other differ significantly. We discuss the effects of the smaller galaxy masses and different mass ratios within our simulations in chapter 6.

For our second SQ model, we adopt the total masses of \citet{HwSt12}. However, most of the other parameters are chosen in accordance with \cite{ReAp10}. Again, our galaxy models consist of a cold dark matter halo, an exponential stellar disc, a stellar bulge and an exponential gaseous disc. Thus, compared to \citet{HwSt12}, our models additionally contain a bulge. We use a high resolution of the galaxy models. As the true nature of NGC 7318a is still unclear, we have modeled it as a disc galaxy. Thus, the parameters of the galaxy models for our two SQ models differ mainly in the mass (and size) of the galaxies. The parameters common to both of our SQ models are given in Table 1.

\begin{table}
\caption{Gravitational softening lengths $\epsilon$.}
\begin{center}
\renewcommand{\arraystretch}{1.2}
\begin{tabular}{lllll}
\hline\hline
&Halo&Disk&Gas&Bulge\\
&[pc/h]&[pc/h]&[pc/h]&[pc/h]\\\hline
SQ model A&113&28&28&28\\
SQ model B&48&10&10&10\\
\hline
\end{tabular}
\end{center}
\label{TS}
\end{table}

For simplicity, we refer to the modified galaxy models with "galaxy model A" and accordingly with "SQ model A" to the total initial SQ setup including all galaxy models and an ambient IGM (see section 3.2) for our improvement based on the Renaud model and with "galaxy model B" and accordingly "SQ model B" to our realization of the Hwang model.

The total masses, virial radii and stellar disc scale lengths and particle numbers used in our galaxy models are summarized in
Table 2. In case of model A, the setup results in particle masses of $m_{\text{gas}}=m_{\text{disc}}=m_{\text{bulge}}\approx2.1\cdot10^{5}$ $h^{-1}$ M$_{\sun}$ and $m_{\text{halo}}\approx2.1\cdot10^{6}$ $h^{-1}$ M$_{\sun}$ with $h = 0.7$. For model B, the particle masses are $m_{\text{gas}}=m_{\text{disc}}=m_{\text{bulge}}\approx9.8\cdot10^{3}$ $h^{-1}$ M$_{\sun}$ and $m_{\text{halo}}\approx9.8\cdot10^{4}$ $h^{-1}$ M$_{\sun}$. We note that these particle masses are of order of the largest molecular clouds, i.e. small-scale shock turbulence is not modeled in our work. The fixed gravitational softening lengths (see e.g. \citealt{De01,JoNa09}) are listed in Table 3. The softening lengths have been adjusted using $\epsilon_{\text{new}} = \epsilon_{\text{old}} \cdot \left[(N^{\text{old}}/{N^{\text{new}}}) \cdot (M_{\text{tot}}^{\text{new}}/ M_{\text{tot}}^{\text{old}})\right] ^{1/3}$. The minimum SPH smoothing length for the gas particles is 1.0 $\epsilon$.

\begin{table*}
 \begin{minipage}{170mm}
\caption{Comparison of initial SQ models.}
\begin{center}
\renewcommand{\arraystretch}{1.2}
\begin{tabular}{lcccc}
\hline\hline
&SQ model of&SQ model A&SQ model of&SQ model B\\
&\citet{ReAp10}&&\citet{HwSt12}&
\\\hline
DM halo&\checkmark&\checkmark&\checkmark&\checkmark\\
Stellar bulge&\checkmark&\checkmark&-&\checkmark\\
Stellar disc&\checkmark&\checkmark&\checkmark&\checkmark\\
Gaseous disc&-&\checkmark&\checkmark&\checkmark\\
Ambient gaseous IGM&-&\checkmark&-&\checkmark\\
NGC 7318a modeled as elliptical&\checkmark&-&\checkmark&-\\
Formation scenario&1&1&3&3\\
Mass ratios of galaxies$^{\text{a}}$&$1:0.2:0.3:0.2$&$1:0.2:0.3:0.2$&$1:0.2:0.7:0.6$&$1:0.2:0.7:0.6$\\
Order of total masses of galaxies&$\approx 10^{11}M_{\sun}$&$\approx 10^{11}M_{\sun}$&$\approx 10^{10}M_{\sun}$&$\approx 10^{10}M_{\sun}$\\
Resolution: mass per star (gas)-particle&$\approx 3\cdot 10^{6}M_{\sun}$&$\approx 3\cdot 10^{5}M_{\sun}$&-&$\approx 1\cdot 10^{4}M_{\sun}$\\
Extension DM halo$^{\text{b,c}}$&46 kpc&280 kpc&135 kpc&102 kpc\\
\hline\hline
\multicolumn{5}{l}{\scriptsize{(a) Mass ratio is given for NGC 7319 : NGC 7320c : NGC 7318a : NGC 7318b.}}\\
\multicolumn{5}{l}{\scriptsize{(b) Cut-off radius for SQ model of \citet{ReAp10} and \citet{HwSt12}, virial radius r200 for SQ model A and B.}}\\
\multicolumn{5}{l}{\scriptsize{(c) The DM halo extensions are given for NGC 7319.}}\\
\end{tabular}
\end{center}
\end{minipage}
\label{cp}
\end{table*}

Finally we want to note that our models differ from the correspondent models of Renaud or Hwang in some details: In our representation of the Renaud galaxy models (galaxy models A), the dark matter haloes of the galaxies are more extended and the disc scale lengths differ from the original model. These differences are on the one hand due to the enhancement of the original setup method for the galaxies \citep{He93} by \cite{SpDi05b}, and on the other hand due to the additionally included gaseous component. We feel confident that our new models are appropriate realisations of the SQ galaxies as they now contain gas and show more realistic extensions of the dark matter haloes. In our representation of the Hwang galaxy models (galaxy models B), the galactic disc sizes differ from the original model and we additionally include a bulge component. As in the restricted three-body simulations of \citet{HwSt12} gas and star disc masses are negligible, we modeled the galaxies for our model B with parameters in analogy to galaxy models A. The parameters which are common to both models are given in Table 1. A detailed comparison of the model properties is listed in Table 4.

\subsection{IGM}

In addition to the modifications described above, an ambient IGM is included. The IGM is set up to be composed of additional gas particles surrounding the galaxies similar to \citet{KoLe11}. We arrange the IGM gas particles on a hexagonal closed-packed lattice. The particle masses of the IGM gas particles are adopted from the respective galaxy models.

The IGM fills a volume of 1000 \texttimes 1000 \texttimes 1000 $h^{-3}$ kpc$^{3}$ and we assume a density of $\rho_{\text{IGM}}=10^{-29}$ g cm$^{-3}$, resulting in particle numbers of $N _{\text{IGM}}$= 496 828 for the SQ model A and $N_{\text{IGM}}$= 2 311 367 for the SQ model B.

The IGM is assumed to be already virialized, whereby the temperature within each model is set to the virial temperature at the virial radius of the largest galaxy model NGC 7319:
\begin{equation}
T_{\text{IGM}}=\frac{2}{3}u_{\text{IGM}}\frac{m_{p} \mu}{k_{B}} = \frac{1}{3}\langle v_{200}^{2}\rangle \frac{m_{p} \mu}{k_{B}} \hspace{1mm} [K] \hspace{1mm},
\end{equation}
with the mean molecular weight for a fully ionized gas of primordial
composition $\mu \approx 0.588$, proton mass $m_{p}$ and Boltzmann constant $k_{B}$. This leads to a temperature of the IGM of $T_{\text{IGM}} \approx 9.4 \cdot 10^{5}$ K for the SQ model A and $T_{\text{IGM}} \approx 1.2 \cdot 10^{5}$ K for the SQ model B.

%\begin{figure}
%\vspace{-0.8cm}
%\begin{center}
%\includegraphics[width=0.9\columnwidth]{./pics/drawing_Galaxy3.pdf}
%\end{center}
%\vspace{-4.8cm}
%\caption{\small{Schematic view of a galaxy model. The arrows indicate the direction and distribution of the initial magnetic field in our simulations. Note that the lengths of the arrows do not represent the magnetic field strength.}}
%\label{galsetup}
%\end{figure}

\subsection{Initial positions and velocities of the galaxies}

\citet{ReAp10} were the first to present possible N-body models of SQ. \citet{HwSt12} resimulated their successful models including a gaseous component. However, their final model represents a different formation scenario than \citet{ReAp10}. 

Our models A and B differ in several features (e.g. gaseous component, IGM, dark matter distribution, see section 3.1) from the original models by \citet{ReAp10} and \citet{HwSt12}, respectively. Therefore, we had to adjust the orbital parameters of the galaxies compared to the original models. We found these new orbital parameters of the galaxies, i.e. new initial velocities in case of SQ model A and initial positions and velocities for SQ model B, by performing more than 100 test simulations to find out the best representation. The orbital parameters of the best representation of SQ for each model are found in Table \ref{TG}. The parameters for the disc orientations are the same as in the original models.

\subsection{Initial magnetic fields}

The initial magnetization of the galactic discs is set up using $B_{\text{x}}=B_{\text{gal,0}}$ and $B_{\text{y}}=B_{\text{z}}=0$ G with the
z-axis being the axis of rotation. This setup ensures that the initial field lies in the plane of the galactic disc. The initial magnetic field strength of the galaxies is assumed to be $B_{\text{gal,0}} = 10^{-9}$ G. This value is by three orders of magnitude smaller than the typical observed galactic magnetic field value \citep[e.g.][]{BeBr96}. For comparison, we also performed simulations with an initial galactic magnetic field strength of $B_{\text{gal,0}} = 10^{-8}$ G and $B_{\text{gal,0}} = 10^{-7}$ G.

The initial magnetic field of the IGM is assumed to be uniform in x-direction with an initial value of $ B_{\text{IGM,0}} = B_{\text{IGM,x}}=10^{-9}$ G and the $x-y$ plane being the orbital plane. In this setup, the IGM magnetic field is naturally also pervading the magnetic field of the galaxies. As the intergalactic magnetic field is usually estimated to be less than $10^{-8}$ G, the assumed initial IGM magnetic field value is already close to the observed value.

%%%%%%%%%%%%%%%%%%%%%%%%%%%%%%%%%%%%%%%%%%%%%%%%%%%%%%%%%%%%%%%%%%%%%%%%%%%%%%%%%%%%%%%%%%%%%%%%%%%%%%%%%%%%%%%%%%%%%%%%%%
%%%%%%%%%%%%%%%%%%%%%%%% Numerical Method %%%%%%%%%%%%%%%%%%%%%%%%%%%%%%%%%%%%%%%%%%%%%%%%%%%%%%%%%%%%%%%%%%%%%%%%%%%%%%%%
%%%%%%%%%%%%%%%%%%%%%%%%%%%%%%%%%%%%%%%%%%%%%%%%%%%%%%%%%%%%%%%%%%%%%%%%%%%%%%%%%%%%%%%%%%%%%%%%%%%%%%%%%%%%%%%%%%%%%%%%%%

\begin{table}
\caption{Parameters for the multi phase model common to all galaxy models.}
\begin{center}
\renewcommand{\arraystretch}{1.2}
\begin{tabular}{lll}
\hline\hline  
\multicolumn{3}{c}{\textsc{Multi Phase Model Parameters}} \\ \hline\hline
Gas consumption timescale&t$_{\text{SFR}}$&8.4 Gyr\\
Mass fraction of massive stars&$\beta_{\text{MP}}$&0.1\\
Evaporation parameter&A$_{0}$&4000\\
Effective supernova temperature&T$_{\text{SN}}$&4 $\cdot$ $10^{8}$ K\\
Temperature of cold clouds&T$_{\text{CC}}$&1000 K\\\hline
\end{tabular}
\end{center}
\label{TM}
\end{table}

\section[]{Simulation Method}

The simulations of SQ presented in this paper are performed with the N-body/SPH code \textsc{Gadget} \citep{SpYo01,Sp05}. Hydrodynamics is included with a formulation of SPH which conserves both energy and entropy \citep{SpHe02}. The simulation results presented in this work were obtained with the development version of \textsc{Gadget}-3, whereby the evolution of magnetic fields can be followed using the additional MHD implementation of \citet{DoSt09}. For a detailed description of SPH and SPMHD methods see e.g. \citet{DoSt09}, \citet{StDo12} and \citet{Pr12}.
For all of the simulations, SPMHD is adopted using the standard (direct) magnetic field implementation, where the magnetic field is evolved using the induction equation. Additionally, within shocks the standard artificial viscosity and the artificial magnetic dissipation are applied. For the dimensionless parameters we use values of $\alpha_{v}=2.0$ (viscosity), $\alpha_{B}=0.5$ (dissipation) and $\beta=1.5$ \citep[e.g.][]{Pr12}. We do not use a subgrid model for physical or turbulent magnetic dissipation.

The Lorentz force, describing the feedback on the plasma caused by the magnetic field, is taken into account as a contribution to the acceleration of each gas particle. However, in cases of strong magnetic forces this momentum conserving form can lead to numerical instabilities, i.e. clumping of the particles \citep{PhMo85}, due to the non-vanishing numerical divergence of the magnetic field, which will not be stabilized when the magnetic pressure exceeds the gas pressure (see e.g. \citealt{DoSt09}, \citealt{StDo12} and \citealt{Pr12}). To avoid these numerical instabilities, the effects of any unphysical source terms of the magnetic field (i.e. any numerically non-vanishing divergence) are subtracted from the momentum equation as suggested by \citet{BoOm01}. A threshold for this divergence force subtraction (half the value of the current Lorentz force) is applied to account for situations in which the acceleration due to the divergence force could become dominant \citep[see][]{StDo12}.

The MHD version of \textsc{Gadget} was already successfully employed for studies of the magnetic field evolution in molecular clouds, in isolated and interacting galaxies as well as in galaxy formation simulations \citep[e.g.][]{BuCl11,KoLe09,GeKo12,BeLe12}.

Radiative cooling, star formation (SF) and associated supernova feedback are applied using the hybrid multiphase model described in \citet{SpHe03}. The interstellar medium is modeled as a multi-phase gas consistent of condensed cold clouds embedded in an ambient hot gas at pressure equilibrium. These cold clouds are forming stars, of which a certain fraction is expected to die instantly as supernovae. Thereby, the feedback energy resulting from supernovae directly heats the ambient hot phase, which in turn looses energy by radiative cooling assuming a primordial gas composition and a zero-metallicity cooling function. The parameters of the multi-phase model used for all of the galaxy models are chosen in accordance with \citet{JoNa09} and are given in Table \ref{TM}. For further details see the corresponding numerical papers on the \textsc{Gadget} code \citep{SpYo01,SpHe02,Sp05}, the SPMHD method \citep[e.g.][]{DoSt09,StDo12,Pr12} or the star formation model \citep{SpHe03}.

%%%%%%%%%%%%%%%%%%%%%%%%%%%%%%%%%%%%%%%%%%%%%%%%%%%%%%%%%%%%%%%%%%%%%%%%%%%%%%%%%%%%%%%%%%%%%%%%%%%%%%%%%%%%%%%%%%%%%%%%%%
%%%%%%%%%%%%%%%%%%%%%%%% Simulation Results %%%%%%%%%%%%%%%%%%%%%%%%%%%%%%%%%%%%%%%%%%%%%%%%%%%%%%%%%%%%%%%%%%%%%%%%%%%%%%
%%%%%%%%%%%%%%%%%%%%%%%%%%%%%%%%%%%%%%%%%%%%%%%%%%%%%%%%%%%%%%%%%%%%%%%%%%%%%%%%%%%%%%%%%%%%%%%%%%%%%%%%%%%%%%%%%%%%%%%%%%

\section[]{Simulations}
Below, we present the results of our simulations of the models A and B (cf. section 3). In all of our simulations, the numerical divergence $\langle h |\nabla \cdot \textbf{B}|/\textbf{B} \rangle$ (cf. Fig. 3) stays essentially below the tolerance value of unity (for a more detailed discussion of the numerical divergence, see e.g. \citealt{KoKa10,BuCl11,BeLe12,GeKo12}), which implies the numerical reliability of the simulations. Unless specified otherwise, all the plots correspond to an initial magnetic field strength of $B_{\text{IGM}}=B_{\text{gal}}=10^{-9}$ G.

\begin{figure}
\begin{center}
\hspace*{-3mm}
\includegraphics[width=1.05\columnwidth]{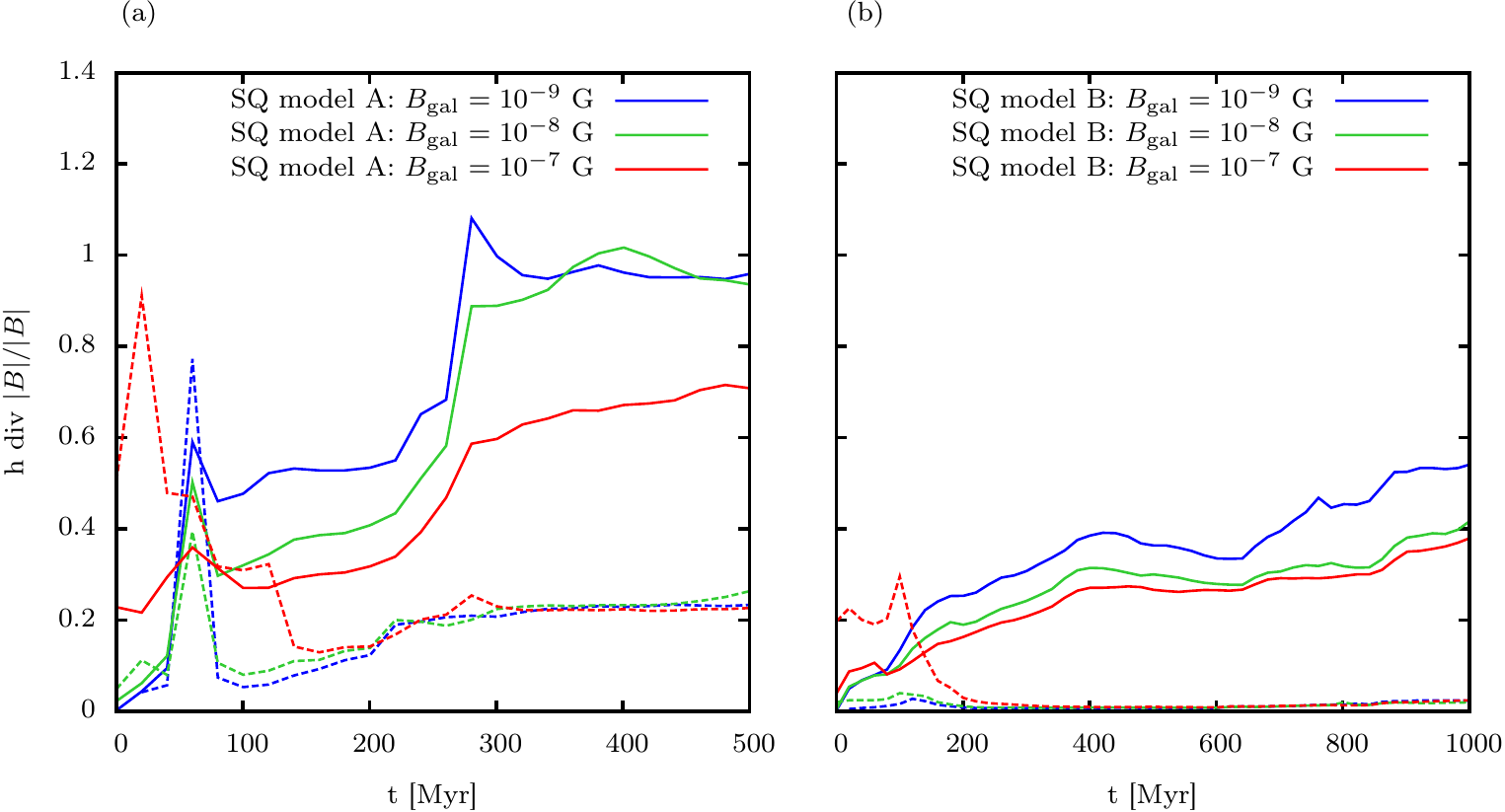}
\caption{\small{Mean numerical divergence $\langle h |\nabla \cdot \textbf{B}|/\textbf{B} \rangle$ as a function of time for (a) SQ model A and (b) SQ model B. Galactic magnetic field (solid lines) and IGM magnetic field (dashed lines) are plotted separately using a threshold of 10$^{-29}$ g cm$^{-3}$. The numerical divergence measure stays essentially below the tolerance value of unity during all of the simulations.}}
\end{center}
\label{numdiv}
\end{figure}

\begin{table}
\caption{Coordinates of the four galaxy models of SQ model A and SQ model B at the present-day configuration, respectively.}
\begin{center}
\renewcommand{\arraystretch}{1.2}
\begin{tabular}{lll}
\hline\hline  
&SQ model A&SQ model B\\
&\small{(x,y,z) in [kpc/h]}&\small{(x,y,z) in [kpc/h]}\\\hline
NGC 7319&(-6.0, 3.0, 10.5)&(5.0, 6.0, -17.0)\\
NGC 7320c&(-103.0, 14.0, -70.0)&(-28.0, 30.0, -31.0)\\
NGC 7318a&(37.0, 11.5, -16.5)&(30.0, -9.0, -84.5)\\
NGC 7318b&(29.0, 11.5, 35.0)&(23.0, -3.5, -16.5)\\\hline
\end{tabular}
\end{center}
\label{CO}
\end{table}

\begin{figure*}
\begin{center}
\includegraphics[width=2.\columnwidth]{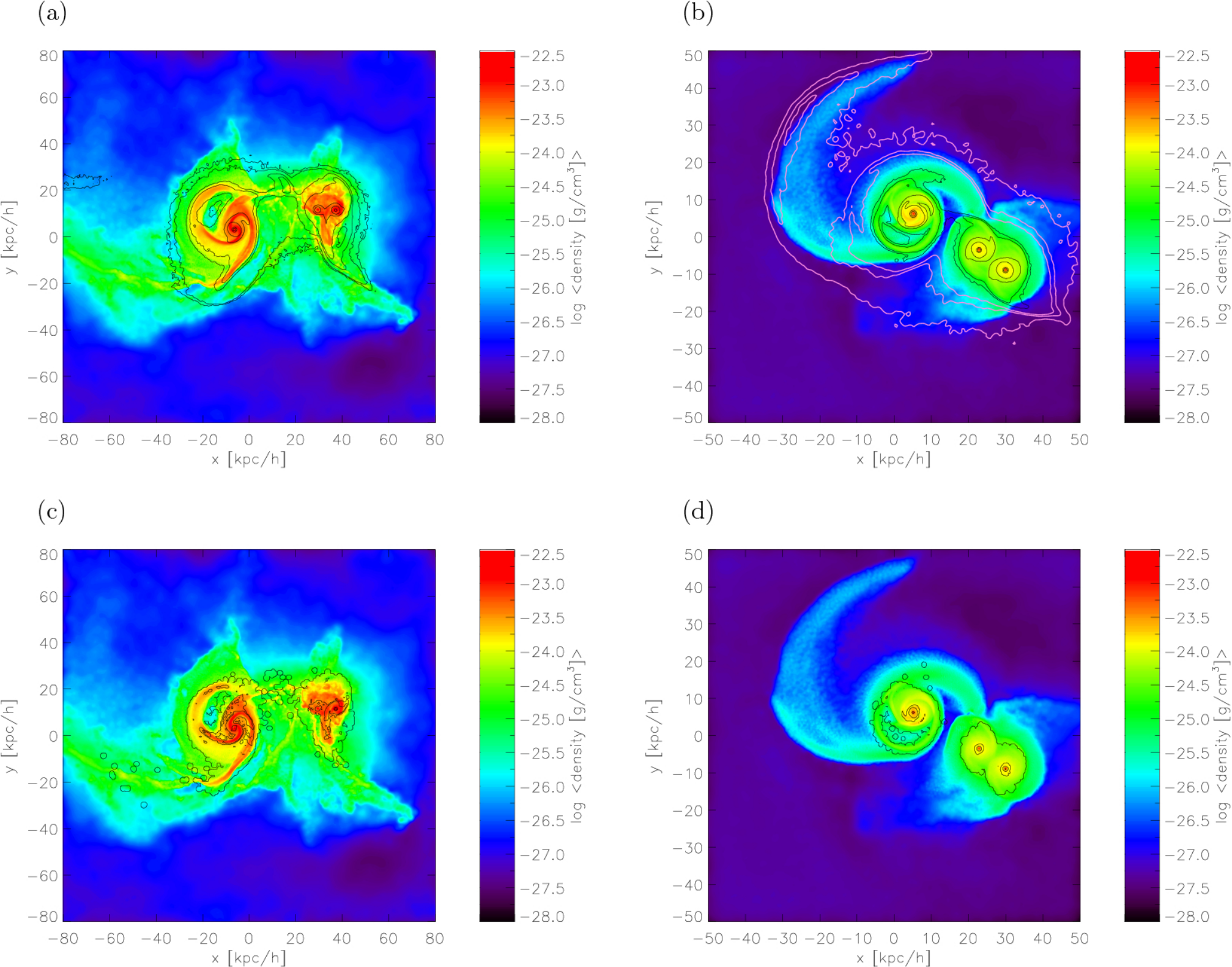}
\end{center}
\caption{\small{(a) Gas density overlaid with contours of the total stellar surface density (contour levels: 0.003, 0.007, 0.03, 0.07, 0.3, 0.7, 3, 7 M$_{\odot}$ pc$^{-2}$) for the present-day configuration of SQ model A ($t_{A3}=320$ Myr). (b) Same as (a), but for the present-day configuration of SQ model B ($t_{B3}=860$ Myr) with three additional pink contour levels (levels: 0.00007, 0.0003, 0.0007 M$_{\odot}$ pc$^{-2}$), which were smoothed with a circular Gaussian beam with FWHM=3. (c) Gas density overlaid with the stellar surface density of new formed stars (contour levels: 3$\cdot 10^{-7}$, 0.003, 0.007, 0.02, 0.07 M$_{\odot}$ pc$^{-2}$) for SQ model A ($t_{A3}=320$ Myr). (d) Same as (c), but for SQ model B ($t_{B3}=860$ Myr).}}
\label{densAB}
\end{figure*}

\subsection[]{General morphology}
In the following, we denote particular evolutionary stages of our simulations with A1 to A3 for the model A, and correspondingly B1 to B3 for model B in order to simplify later reference. 

For SQ model A, it takes 320 Myr from the initial configuration (see Tables 1 and 2) to reach its best fit with observations. In the course of the evolution, first NGC 7320c undergoes a collision with NGC 7319 (A1: $t_{ot}=t_{A1}=80$ Myr), producing the outer tail. After $t_{it}=140$ Myr, NGC 7318a starts interacting with the already disturbed galaxy NGC 7319, resulting in the formation of the inner tail. Afterwards, the final encounter of the discs of NGC 7318a and NGC 7319 takes place at $t_{A2}=240$ Myr. Subsequently, the high-speed intruder NGC 7318b hits the system, which leads about 40 Myr later to a configuration consistent with the morphology of the observed system (A3 - present day: $t_{pd}=t_{A3}=320$ Myr) (cf. Fig. \ref{morph}). The resulting coordinates of the four galaxies at the present-day configuration are listed in Table \ref{CO}.

Fig. \ref{densAB} (a) shows the gas density overlaid with contours of the stellar surface density for SQ model A at the time $t_{A3}=320$ Myr (A3). Only NGC 7319 and the galaxy pair NGC 7318a/b are visible. NGC 7320c lies outside the plotted region (cf. Table \ref{CO}). Qualitatively, the model shows a good agreement of the general features with observations, i.e. the morphology of the large galaxy NGC 7319, the inner tail south-east of NGC 7319, and the structure of the western galaxy pair NGC 7318a/b. Some kind of outflow is indicated in the gas density distribution north of NGC 7319 and also north and south-west of the pair NGC 7318a/b. The stellar surface density shows the highest values within the inner disc of NGC 7319 and also within the disc of NGC 7318a. The stellar density in the disc of NGC 7318b is slightly lower. A bridge between NGC 7319 and the colliding pair NGC 7318a/b is clearly visible in the stellar surface density as well as in the gas density. At the left edge of the plot, there is an elongated region with a stellar surface density larger than 0.003 M$_{\odot}$ pc$^{-2}$. This region belongs to the galaxy NGC 7320c which lies outside the plot. However, the model does not reproduce the observed position of the galaxy pair NGC 7318a/b correctly and the outer tail is generated in this model but is already too diffuse to be visible at the present-day configuration. These aspects were both already noted for the original model by \citet{ReAp10}. Furthermore, the small-scale details of this galaxy pair, e.g. the structure of the spiral arms of NGC 7318b, are not well reproduced.

For our SQ model B it takes 860 Myr from the initial conditions (cf. Tables 1 and 2) to develop a morphology similar to the observed configuration. However, the interaction history differs significantly from model A: At first, a close passage of NGC 7320c around NGC 7319 (B1: $t_{ot}=t_{it}=t_{B1}=160$ Myr) simultaneously forms both the inner and outer tails. Subsequently, the galaxies NGC 7318a and NGC 7318b undergo a collision behind the orbital plane of the main system consisting of the galaxies NGC 7319 and NGC 7320c (B2: $t_{B2}=640$ Myr). Afterwards, the high-speed intruder NGC 7318b is moving towards the main system and collides with the IGM material west of NGC 7319. Meanwhile, about 80 $h^{-1}$kpc behind the plane of the main system, NGC 7318a moves westwards. The evolution results in a configuration similar to observations (B3 - present day: $t_{pd}=t_{B3}=860$ Myr). The resulting coordinates of the four galaxies at the present-day configuration are listed in Table \ref{CO}. Within this model, NGC 7320c is still present in the tail, but however, it is largely disrupted, therefore we had to estimate the coordinates of the center of this galaxy.

\begin{figure}	
\includegraphics[width=1.\columnwidth]{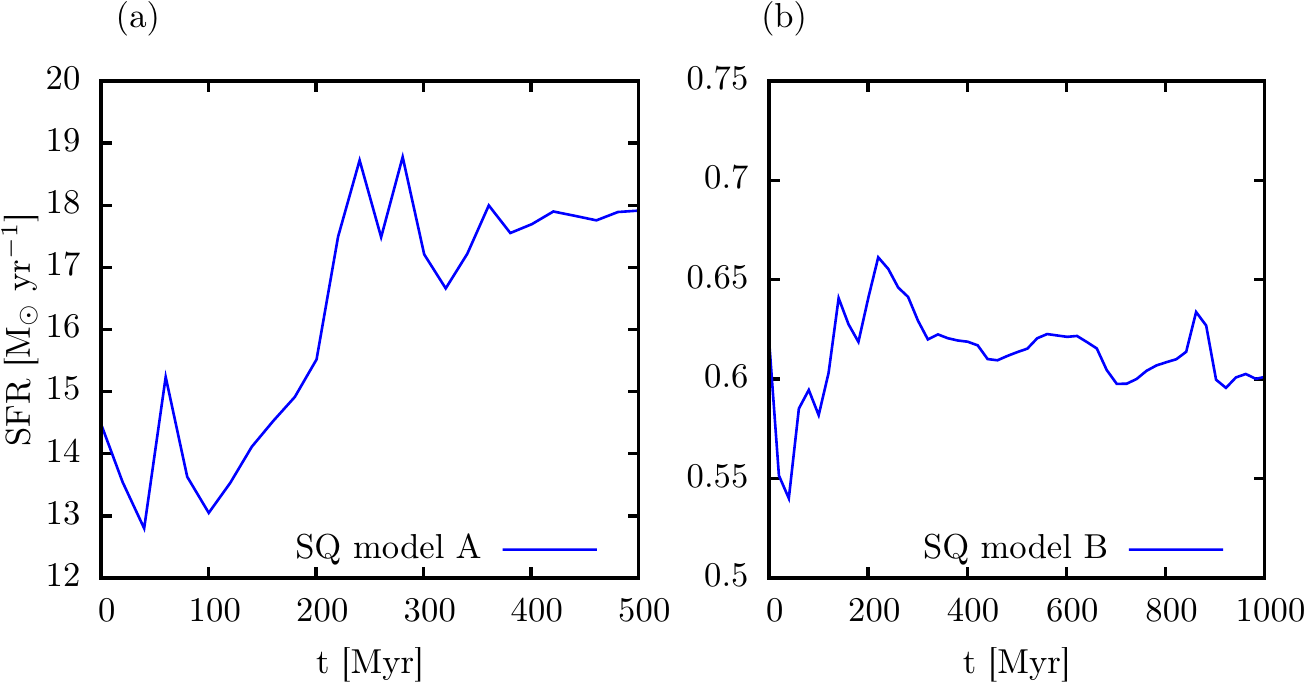}
\caption{\small{SFR as a function of time for (a) SQ model A and (b) SQ model B. Note the different scaling of the axes. }}
\label{sfr}
\end{figure}

\begin{figure*}
\begin{center}
\includegraphics[width=2.\columnwidth]{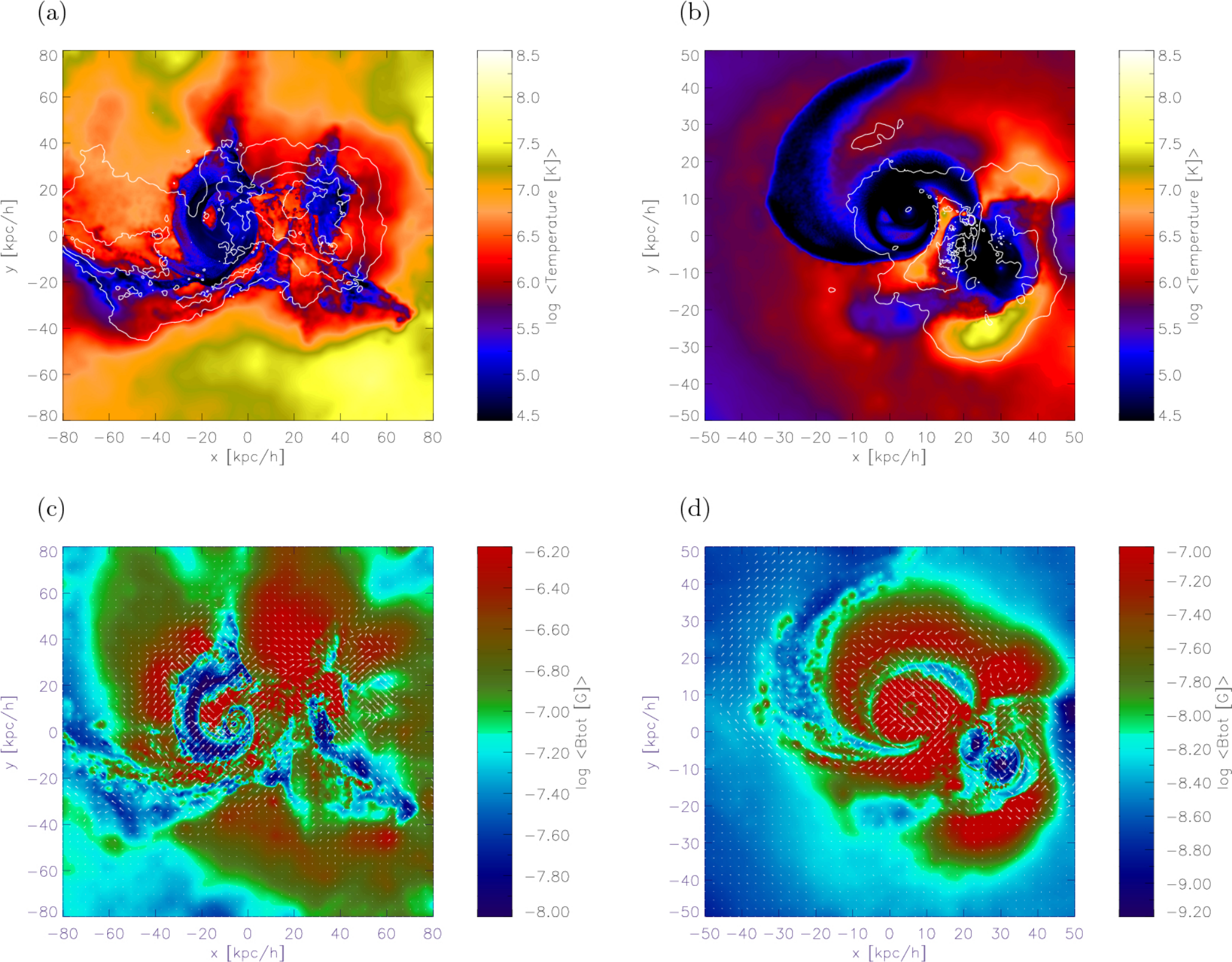}
\end{center}
\caption{\small{(a) Temperature overlaid with X-ray contours (logarithmic contour levels: 36.75, 37.25, 37.75 and 38.25 erg s$^{-1}$) for the present-day configuration of SQ model A at the time $t_{A3}=320$ Myr. (b) Same as (a), but for the present-day configuration of SQ model B ($t_{B3}=860$ Myr) and with lower contour levels (logarithmic contour levels: 35.75,36.25,36.75,37.25 erg s$^{-1}$). (c) Magnetic field strength overlaid with arrows showing the direction of the magnetic field (initial magnetic field in x-direction with $B_{\text{gal}}=B_{\text{IGM}}=10^{-9}$ G) for SQ model A ($t_{A3}=320$ Myr). (d) Same as (c), but for SQ model B ($t_{B3}=860$ Myr).}}
\label{tempAB}
\end{figure*}

Fig. \ref{densAB} (b) shows the same quantities as Fig. \ref{densAB} (a) but for the SQ model B at the time $t_{B3}=860$ Myr (B3). As for the model A, there is a good qualitative agreement of the general features with observations. Yet, model B reproduces the position of the galaxy pair NGC 7318a/b better than model A (cf. also Fig 1). The stellar surface density shows the highest values in the inner discs of NGC 7319 and NGC 7318a/b. Again, a stellar bridge is visible between NGC 7319 and the galaxy pair NGC 7318a/b. However, the position of NGC 7318a is slightly too much south, which is a feature of our SQ model B, not of the original model by \citealt{HwSt12}. SQ model B is also not capable of reproducing small-scale features like the spiral arms of NGC 7318b or the detailed structure of NGC 7319 correctly and the outer tail is shorter compared to observations.

SQ model A and SQ model B differ significantly in the formation scenario of the outer and the inner tails. Whereas in the first case the two tails are evolved in two different interactions of NGC 7319 with NGC 7320c and NGC 7318a, within the SQ model B the tails are created by only one interaction event of NGC 7319 with NGC 7320c. Within SQ model A, the outer tail is formed about $t_{pd}-t_{ot} = 240$ Myr ago and the inner tail about $t_{pd}-t_{it} = 180$ Myr ago, resulting in a formation age difference of $\sim 60$ Myr (which is similar to the results of the model of \citealt{ReAp10}, who found a formation age difference between the tails of $\sim 70$ Myr). In contrast, SQ model B shows an equal formation age of both tails about $t_{pd}-t_{ot} = 700$ Myr ago. However, observations are still suggesting different ages of the tails: \citet{MoSu97} proposed an age of the outer tail of $\geq 500 - 700$ Myr and for the inner tail $\sim 200$ Myr by considering the radial velocity difference between NGC 7319 and NGC 7320c. Later, \citet{SuRo01} found that this measurement of the radial velocity was highly overestimated and suggested a much slower radial velocity for NGC 7320c (almost identical to that of NGC 7319) resulting in a prediction for the encounter of NGC 7319 with 7320c (causing the inner tail) within the second formation scenario (see chapter 2) about $\geq 500$ Myr ago, which is similar to the predicted age of the outer tail. \citet{FeGa11} suggested an age  of $\sim 400$ Myr for the outer tail and $\leq 200$ Myr for the inner tail. However, \citet{FeGa11} also found that the inner tail also contains some old clusters with an age of $\sim 500$ Myr, even if it is mainly composed of blue clusters. \citet{HwSt12} argue that the different formation ages would not have necessarily taken place in order to explain the different star formation histories, as the outer tail evolves in a different environment, which may be less dense compared to the environment of the inner tail.

SQ model B clearly supports the idea of a common origin of the both tails caused by only one interaction event about $\sim 700$ Myr ago, which agrees well with observational predictions for the formation age of the outer tail (e.g. \citealt{MoSu97}). Contrary, SQ model A implies different formation ages of the tails, whereby the finding of a formation age of the inner tail of $\sim 200$ Myr corresponds also to observational findings \citep[e.g.][]{MoSu97,FeGa11}, whereas the formation age difference of only $\sim 60$ Myr seems to be too small compared to observations. To conclude, we cannot rule out any of the proposed formation scenarios and ages for the two tails, however, the slightly better agreement of the resulting positions of the galaxies (particularly of the galaxy pair NGC 7318a/b) and the larger formation age of the tails may be interpreted as an indication to prefer the formation scenario of SQ model B.

\subsection[]{Star formation}

\begin{figure}
\includegraphics[width=1.\columnwidth]{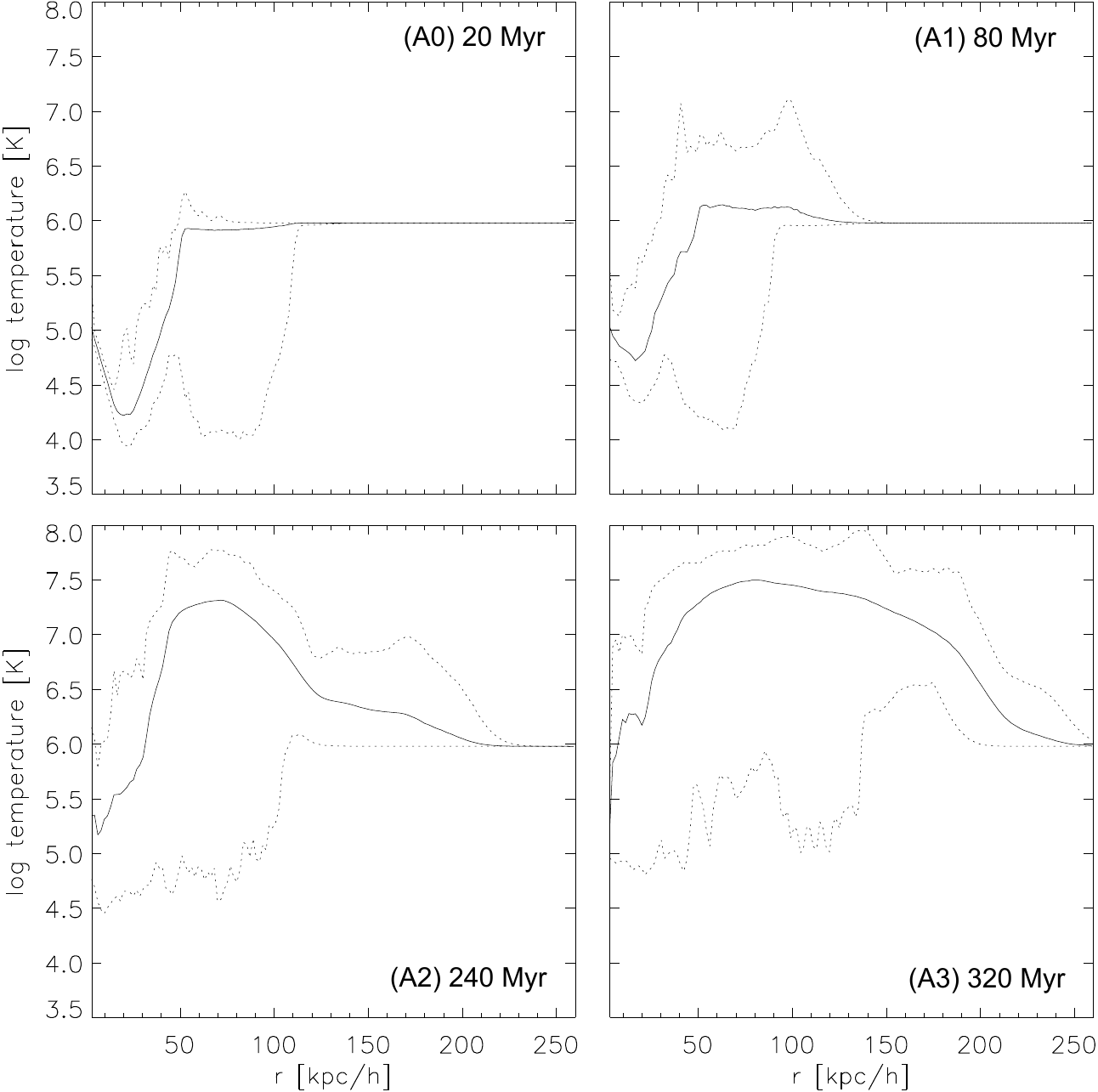}
\caption{\small{Radial profiles of mean (solid lines), maximum (upper dotted lines) and minimum (lower dotted lines) temperatures for SQ model A at different evolutionary stages. Here, the origin is the centre of the main galaxy NGC 7319.  }}
\label{temp1}
\end{figure}

\begin{figure}	
\includegraphics[width=1.\columnwidth]{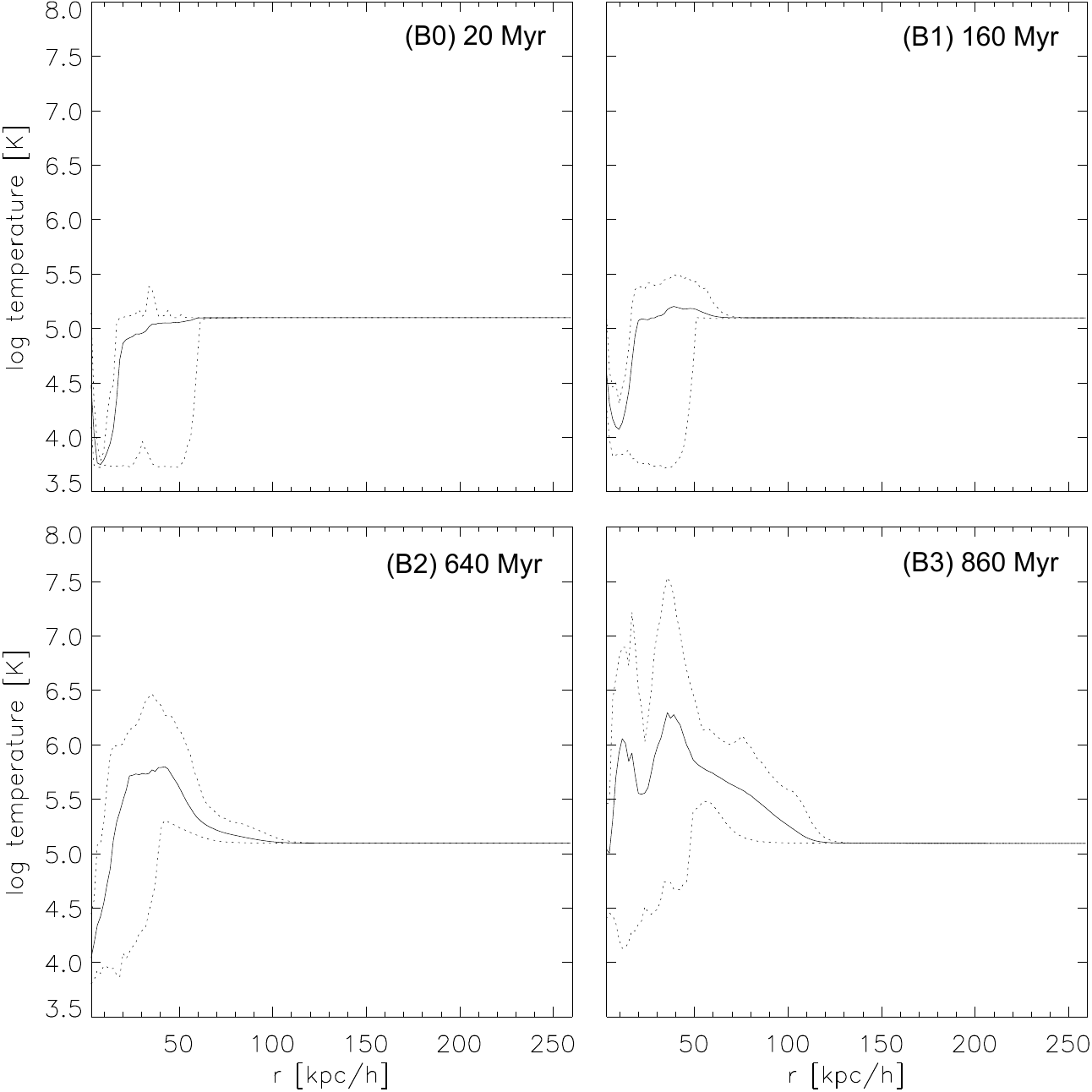}
\caption{\small{Same as Fig. \ref{temp1}, but for SQ model B. The temperature gets enhanced by shocks and outflows, whereas for the more massive galaxies of SQ model A this enhancement is more efficient.}}
\label{temp2}
\end{figure}

Star formation (SF) activity in SQ is believed to be triggered by the interactions \citep{Xu05}. Fig. \ref{densAB} (c) again shows the gas density for model A at $t_{A3} = 320$ Myr, but this time overlaid with contours of the stellar surface density of the new formed stars (i.e. stars which were not present in the initial setup). Most new formed stars are found within the inner disc regions of NGC 7319 and NGC 7318a. Slightly less SF takes place within NGC 7318b. A low star formation activity and thus a low surface density of newly formed stars is found within the spiral arms of NGC 7319, the outer discs of NGC 7318a/b, in the IGM between NGC 7319 and NGC 7318a/b, and in the region of the inner tail south-east of NGC 7319. The surface density of new formed stars within the IGM mainly traces the stellar and gaseous bridge. However, parts of this star-forming region seem to coincide also with probably the edges of a shock region visible in X-ray and radio emission (cf. section 5.3 and 5.5), which in principle corresponds to the observed star-forming regions \citep{Xu05,Cl10} (cf. section 2). However, the starburst region north of NGC 7318a/b found in observations \citep{Xu05} is not revealed in the surface density of new formed stars in our SQ model A. 

Fig. \ref{densAB} (d) shows the same quantities as Fig. \ref{densAB} (c) but for the present-day configuration of SQ model B. The highest surface density of new formed stars is found within the inner discs of NGC 7319 and NGC 7318a/b. Less star formation takes place in the outer discs of the galaxies and within the spiral arm structure of NGC 7319. The lowest star formation is found north of NGC 7319. Most of these regions are indeed also observed to form stars \citep{Xu05} (cf. section 2). However, there is no region of noticeable ongoing star formation in the IGM between the main galaxy NGC 7319 and the galaxy pair NGC 7318a/b, i.e within or at the edges of the supposed shock region. The starburst region north of the pair NGC 7318a/b found in observations \citep{Xu05} is also not reproduced.

The SF rate (SFR) of isolated galaxies is found to be approximately constant, whereby the constant SFR depends on the total masses and gas fractions of the progenitors \citep{Co04}. This behaviour results in significantly higher starting values of the SFR in our SQ model A compared to SQ model B. Galaxy interactions trigger new SF as they efficiently compress the gas and therefore lead to a conversion of cold gas into new stars (cf. section 4). The ability of the system to trigger prominent star bursts depends on the initial masses, mass ratios and on the orbit of the progenitor galaxies \citep{Co04}. Fig. \ref{sfr} shows the total SFR as a function of the time for SQ model A (a) and SQ model B (b), respectively. In good agreement with previous studies of SFR during galaxy minor mergers \citep{Co04}, the SFR strongly depends on the initial masses of the progenitor galaxies, resulting in a significantly higher SFR for SQ model A compared to SQ model B. In both simulations, the interactions significantly enhance the overall SFR, whereby the more massive galaxies of SQ model A lead to stronger star bursts compared to SQ model B.

\subsection[]{Temperature and X-ray emission}

The main source of X-ray emission is hot gas, which is heated by shocks accompanying the interactions. 
We calculate the bolometric X-ray luminosity following the method of \citet{NaFr95}, which assumes thermal bremsstrahlung to be the main X-ray source in agreement with the applied zero-metallicity cooling function. The bolometric X-ray luminosity is projected along the line of sight according to
\begin{equation}
L_{\text{x}} = 1.2 \cdot 10^{-24} \frac{1}{\left(\mu m_{\text{p}}\right)^{2}} \hspace{1mm} \sum_{i=1}^{N_{\text{gas}}} m_{\text{gas},i} \hspace{1mm} \rho_{i} \left( \frac{k_{\text{B}} T_{i}}{\text{keV}}\right)^{1/2} \left[\frac{\text{erg}}{\text{s}} \right],
\end{equation}
with mass $m_{\text{gas},i}$, density $\rho_{i}$ and temperature $T_{i}$ of the $i$-th gas particle in cgs-units, respectively. Only fully ionized particles should be considered when calculating the luminosity. Therefore, we exclude contributions of particles with temperatures lower than $10^{5.2}$ K and densities higher than $0.01 \text{M}_{\odot} \text{pc}^{-3}$ \citep[cf.][]{CoDi06}. 

Fig. \ref{tempAB} (a) shows the temperature overlaid with contours of the X-ray emission for the present-day configuration of SQ model A. The gas within the galaxies is cooler, whereas the IGM gas is heated by shocks and outflows caused by the interactions, which is in good agreement with previous studies \citep[e.g.][]{KoLe11,GeKo12}. The logarithmic contours illustrating the X-ray emission reveal a total X-ray luminosity which is approximately four orders of magnitude lower than the observed X-ray luminosity in SQ ($10^{40} - 10^{41}$ erg s$^{-1}$, \citealt{SuRo01}). This low  X-ray luminosity results most probably from the applied zero-metallicity cooling and the lack of black holes in the simulations (cf. \citealt{CoDi06,GeKo12}). Nevertheless, the X-ray luminosity shows the highest values in the IGM region between NGC 7319 and NGC 7318b, indicating a large shock east of NGC 7318b. This shock region fits very well to the observed shock front visible as a ridge in the X-ray and radio emission (cf. section 2 and Fig. \ref{Xu}). However, the morphology of the shock region found in our simulations differs slightly from the observations. This difference might be explained by the more northern position of the galaxy pair NGC 7318a/b compared to the observed position.

\begin{figure*}	
\includegraphics[width=1.4\columnwidth]{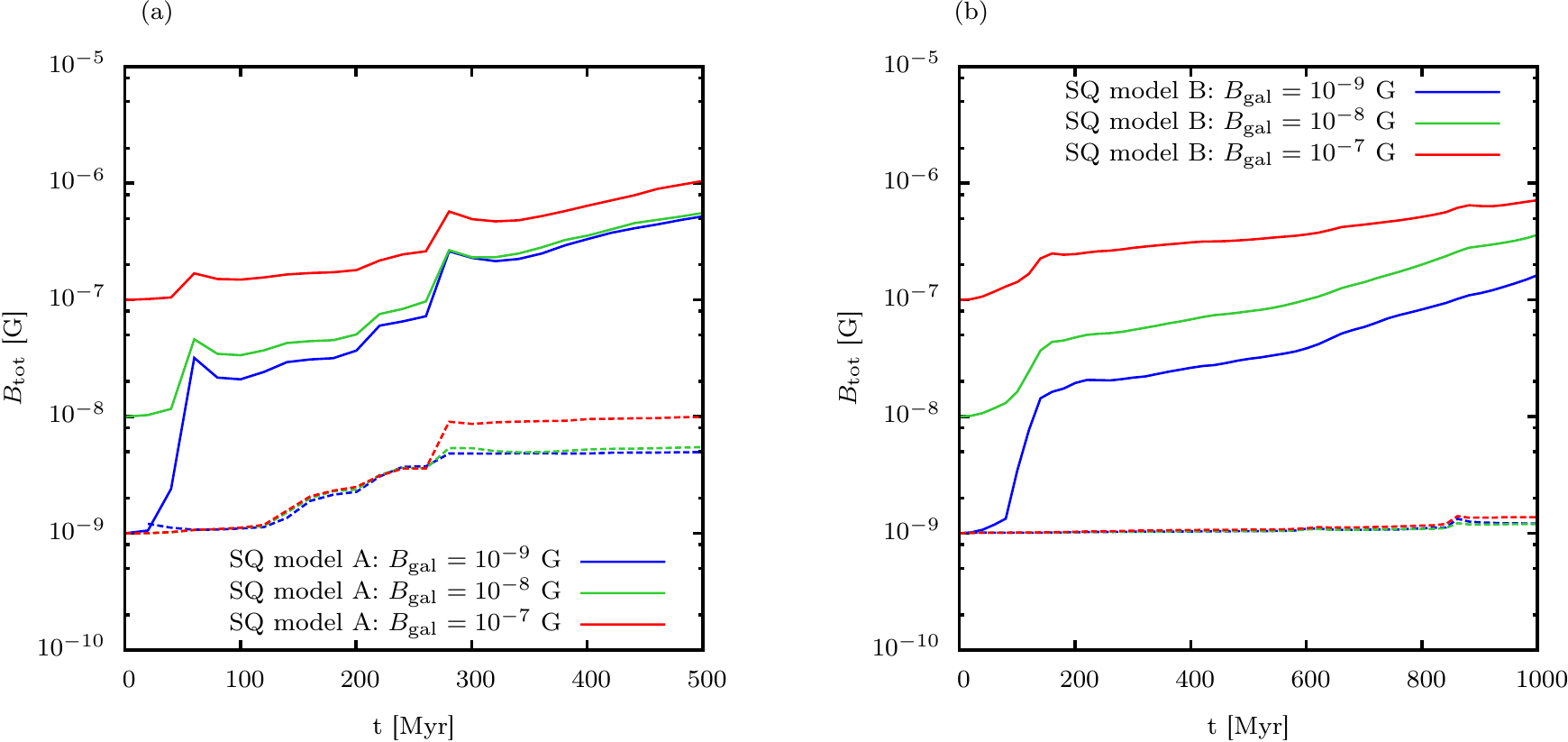}
\caption{\small{Mean total magnetic field strength as a function of time for (a) SQ model A and (b) SQ model B with initial magnetic fields of $B_{\text{IGM}}=10^{-9}$ G for each run and $B_{\text{gal}}=10^{-9}$ G (blue), $B_{\text{gal}}=10^{-8}$ G (green) and $B_{\text{gal}}=10^{-7}$ G (red), respectively. Galactic magnetic field (solid lines) and IGM magnetic field (dashed lines) are plotted separately using a threshold of 10$^{-29}$ g cm$^{-3}$. In all of the simulations, the galactic magnetic field gets efficiently amplified, whereby the maximum magnetic field strengths during the interactions are higher for the more massive galaxies of SQ model A and for higher initial magnetic field strengths.  }}
\label{magfield}
\end{figure*}

Fig. \ref{tempAB} (b) shows the same quantities as Fig. \ref{tempAB} (a) for the present-day configuration of SQ model B. Again, the IGM surrounding the galaxies is heated by the shocks and outflows caused by the interactions. One large outflow is clearly visible in the temperature in a region south of the galaxy pair NGC 7318a/b and a smaller outflow in the north-west of the pair. The gas within the galaxies is cooler than within the IGM. The overall temperature of this model is significantly lower compared to SQ model A (Figs. \ref{tempAB} (a) and \ref{tempAB} (b) use the same colour bar). This is consistent with the smaller and less massive galaxy models. Consequently, also the X-ray emission is approximately one order of magnitude smaller than for SQ model A. As before, the highest X-ray emission is found within the IGM region between NGC 7319 and NGC 7318b, again indicating a prominent shock. This region of enhanced X-ray emission is more extended compared to SQ model A, and agrees in general well with observations (cf. Fig. \ref{Xu}). 

The radial profiles (with the origin being the centre of the main galaxy NGC 7319) of the mean, minimum and maximum temperature within 260 Mpc at different evolutionary stages are shown in Fig. \ref{temp1} for SQ model A and in Fig. \ref{temp2} for SQ model B. In both figures, the upper left panels show the profiles before any collision events at $t=$ 20 Myr (A0 and B0, respectively). A1, A2, A3 and B1, B2, B3, respectively, denote the characteristic interaction events until the present-day configuration is reached (cf. section 5.1). For both models, at $t=$20 Myr (A0 and B0), the temperature is lower at small radii, reflecting to the cooler gas within the galaxies. At greater radii, the mean temperature is constant. This is due to the initial setup, where the temperature was assumed to be already virialized (see section 3.2). Within the inner 50 - 100 $h^{-1}$ kpc the small minimum values of the temperatures correspond to the gas within the smaller galaxies.
In both models, the temperature gets enhanced by shocks and outflows caused by the interactions (A1-A3 and B1-B3, respectively). The increase in temperature is propagating towards higher radii due to the dilatation of the shock-heated regions (A2, A3 and B2, B3, respectively). The gas inside the galaxies is also heated by the interactions but it cools down again between the interactions. Consequently, the overall temperature inside the galaxies is generally much lower than the temperature of the IGM gas. As already described above, the larger and more massive galaxies of SQ model A enhance the IGM temperature more efficiently (Fig. \ref{temp1}) compared to SQ model B (Fig. \ref{temp2}), in agreement with previous studies \citep{GeKo12}. Moreover, the propagation of the shock-heated regions within the IGM is much more efficient for the SQ model A, because the smaller galaxies of SQ model B cause weaker shocks and thus a slower propagation of the shock-heated regions.

\subsection[]{Magnetic field structure}

The magnetic field is expected to get enhanced through random and turbulent motions driven by the interactions of the galaxies \citep[see e.g.][for a review]{BrSu05}. The compact SQ system has undergone a number of interactions, whereby the magnetic field should have been amplified significantly. Based on observational evidence, \citet{XuLu03} suggests that the shock front revealed by the X-ray and radio emission contains ionized gas and cold dust as well as hot thermal electrons, relativistic electrons and magnetic fields.
This assumption agrees very well with expectations motivated by simulations: interaction-driven shocks are propagating favorably into the IGM, thereby heating the IGM gas and thus producing hot thermal electrons \citep{KoLe11}. Also, material of the galactic disc, i.e. ionized gas, is transported with the shock. Within the shock, electrons could become accelerated to relativistic velocities via the Fermi-acceleration process, in turn enhancing the radio emission in these regions. The IGM magnetic field is amplified by shocks due to the compression of the field lines in front of the shock and turbulence behind the shock front.

Fig. \ref{tempAB} (c) shows the mean total magnetic field strength overlaid with the magnetic field vectors for the SQ model A. It can be recognized that shocks and interaction-driven outflows are expanding into the IGM, thereby enhancing the magnetic field strength up to values of almost $\mu$G order. Some regions, such as the upper galactic arm of NGC 7319, the region around the inner tail, and parts within the galaxy pair NGC 7318a/b show lower values of the total magnetic field strength. These regions are most probably not directly affected by the encounters. Hence, the slight enhancement of the magnetic field in these regions might be solely due to the winding process of the galaxies. High values of the magnetic field strength are in particular found within the large outflow in the upper north which seems to originate between NGC 7319 and NGC 7318b, where also the large shock is located. There, we also find high magnetic field strengths. The direction of the magnetic field, which is heading towards NGC 7318b, may indicate that the shock was triggered by a collision between NGC 7318b with the IGM (instead of a collision with NGC 7318a), thus supporting the common hypothesis of the origin of the shock (cf. section 2).

\begin{figure}
\includegraphics[width=1.\columnwidth]{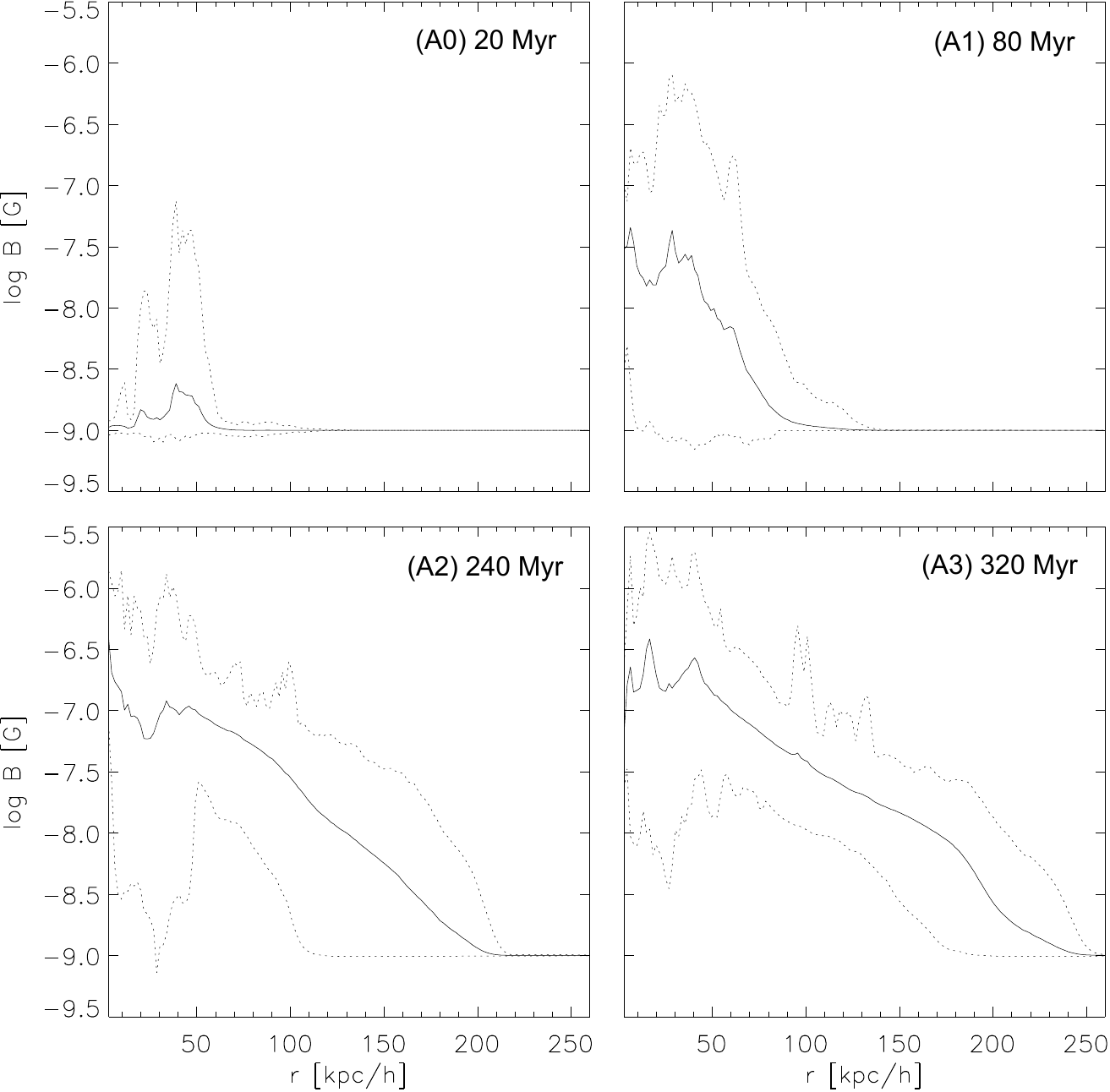}
\caption{\small{Radial profiles of mean (solid lines), maximum (upper dotted lines) and minimum (lower dotted lines) magnetic field strengths (with the origin being the centre of the main galaxy NGC 7319) for SQ model A at different evolutionary stages.  }}
\label{magn1}
\vspace{0.01cm}
\includegraphics[width=1.\columnwidth]{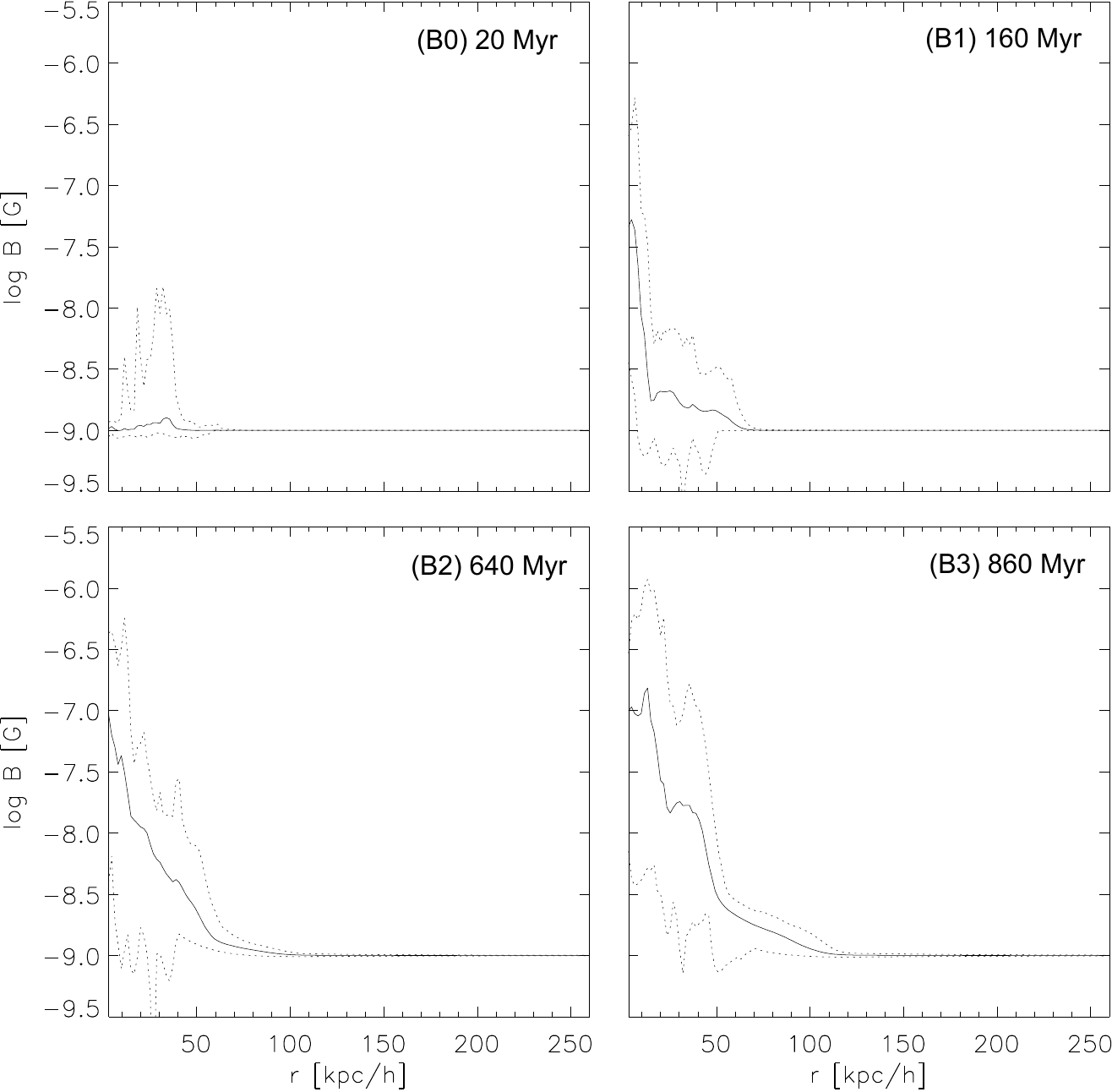}
\caption{\small{Same as Fig. 10, but for SQ model B. In both models, the magnetic field strength within the galaxies gets efficiently amplified during the interactions. The amplification is more efficient for SQ model A. }}
\label{magn2}
\end{figure}

Fig. \ref{tempAB} (d) shows the same quantities as Fig. \ref{tempAB} (c) but for the SQ model B. Note the different color bar appropriate for the lower magnetic field strengths. As before, shocks and interaction-driven outflows are propagating into the IGM, enhancing the magnetic field strength up to values of order 0.1 $\mu$G. Again, regions of lower magnetic field strength are found between the tidal arms of NGC 7319 and within the two galaxies NGC 7318a/b. These regions are probably not directly affected by the interactions (the galaxies NGC 7318a/b are affected by their mutual collision only in the very outer regions of the discs). The highest values of the magnetic field strength are found between the main galaxy and the galaxy pair NGC 7318a/b. Again, this highly magnetized region coincides with the supposed shock region also observed in X-ray. Comparable magnetic field strengths are found within the disc of NGC 7319 and within the outflows north of NGC 7319 and south-west of NGC 7318a/b. The magnetic field vectors within the region of the shock are either directed towards NGC 7318b or along the inner tail. As the collision of NGC 7318b with NGC 7318a happened about 220 Myr before the present-day configuration, the prominent shock found in SQ model B is certainly a result of an interaction of NGC 7318b with the IGM.

\begin{figure*}
	\begin{center}
\includegraphics[width=1.5\columnwidth]{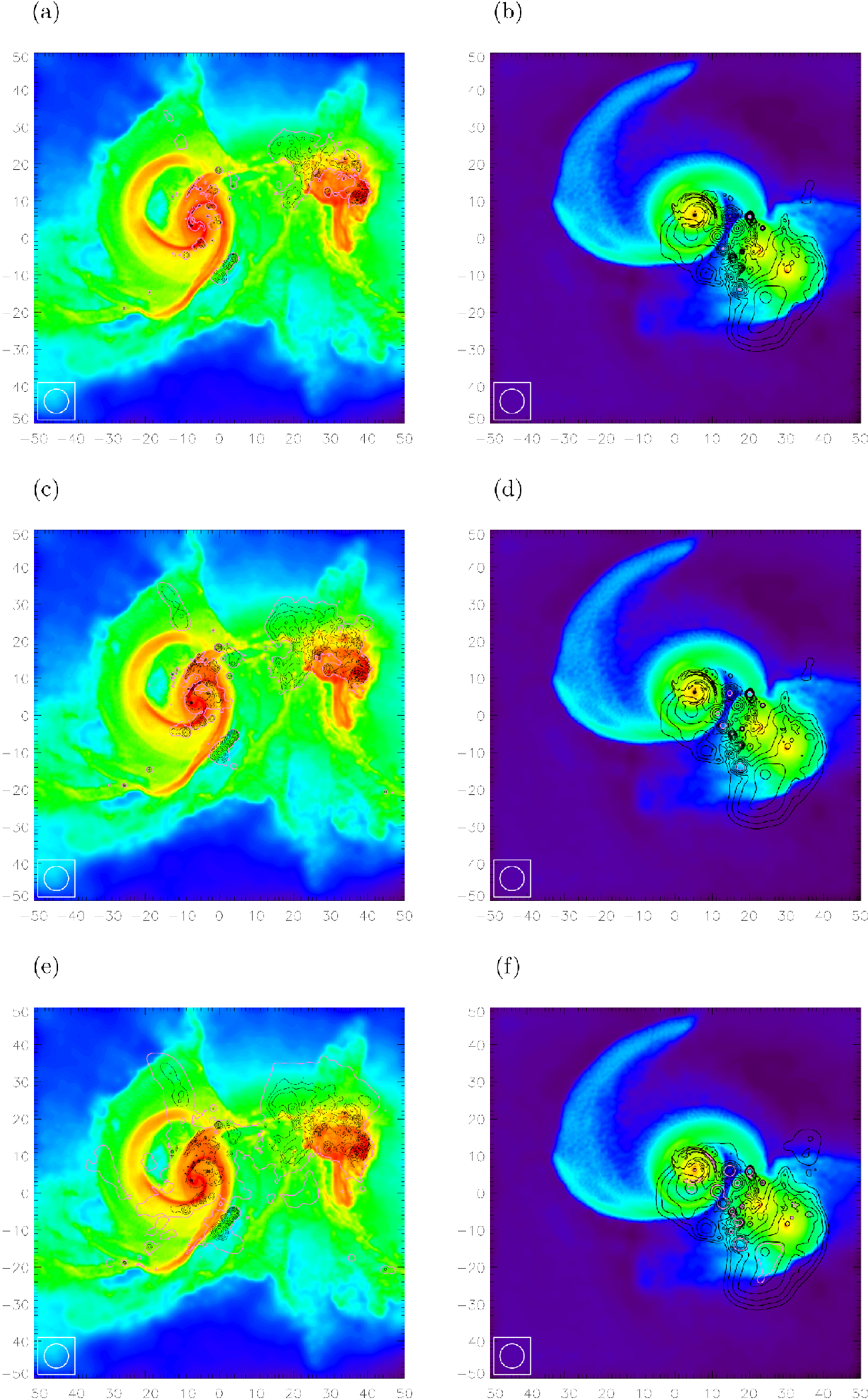}
	\end{center}
\caption{\small{ Radio maps showing the total Intensity $I_{tot}$ of model A (left panels) and model B (right panels), respectively, at 4.86 GHz (upper panels), 1.4 GHz (middle panels) and 0.2 GHz (lower panels). Colors visualize the gas density (color coding same as in Fig. \ref{densAB}a (left) and Fig. \ref{densAB}b (right). Upper panels: (a) the contour levels are 50, 100, 200, 400, 800, 1600 $\mu$Jy/beam and (b) 0.5, 1, 2, 4, 8, 16 $\mu$Jy/beam. Middle panels: (c) the contour levels are 90, 180, 360, 720, 1440, 2880 $\mu$Jy/beam and (d) 1.2, 2.5, 5, 10, 20, 40 $\mu$Jy/beam. Lower panels: (e) the contour levels are 300, 600, 1200, 2400, 4800, 9600 $\mu$Jy/beam and (f) 2.5, 5, 10, 20, 40, 80 $\mu$Jy/beam. The pink contour in each plot corresponds to 50 $\mu$Jy/beam. We assume a beamsize of $20''$. Within both models, the pink contour level encloses larger areas for lower frequencies.}}
\label{radiotot}
\end{figure*}

\begin{figure*}
	\begin{center}
\includegraphics[width=2.0\columnwidth]{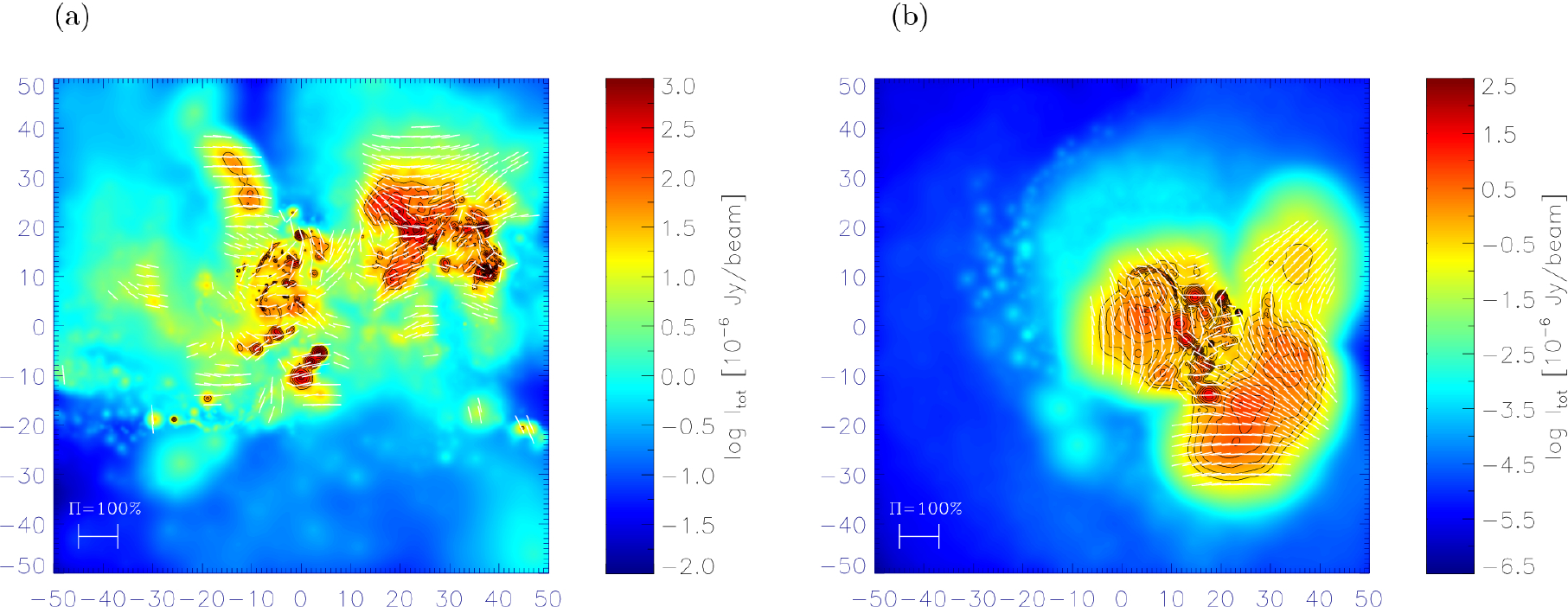}
	\end{center}
\caption{\small{ Synthetic radio map for model A (a) and model B (b), respectively, at 4.86 GHz. Colors visualize the total intensity $I_{tot}$ (in $\mu$Jy/beam). Black contours show the polarized Intensity $I_{p}$ and correspond to (a) to 25, 50, 100, 200, 400, 800, 1600 $\mu$Jy/beam and (b) to 0.25, 0.5, 1, 2, 4, 8, 16 $\mu$Jy/beam. Magnetic field lines derived from calculations of polarization are shown in white. Note the different color coding of the radio maps. Within both models, high total and polarized synchrotron emission is found within the region of the prominent shock between NGC 7319 and NGC 7318a/b. }}
\label{radiopol}
\end{figure*}

Fig. \ref{magfield} shows the evolution of the mean total magnetic field strength as a function of time for the SQ model A (a) and B (b) with initial magnetic fields of $B_{\text{IGM}}=10^{-9}$ G within each run and $B_{\text{gal}}=10^{-9}$ G (blue lines), $B_{\text{gal}}=10^{-8}$ G (green lines) and $B_{\text{gal}}=10^{-7}$ G (red lines), respectively. We plot the galactic magnetic field (solid lines) and the IGM magnetic field (dashed lines) separately using a threshold of 10$^{-29}$ g cm$^{-3}$. The presented simulations show a slight amplification of the total galactic magnetic field before the first encounter due to the winding process, which is in good agreement with previous studies \citep[e.g.][]{KoLe09,KoKa10}. During the first interaction of NGC 7319 with NGC 7320c ($t_{A1}=80$ Myr and $t_{B1}=160$ Myr, respectively) the galactic magnetic field gets efficiently amplified, whereby the maximum magnetic field strength reached during this interaction are higher for the more massive progenitor galaxies of SQ model A. Also, they are higher for higher initial magnetic field strengths, in agreement with previous simulations \citep{GeKo12}. Within all of the simulations, further interactions of the galaxies (A2 and A3, B2 and B3, cf. section 5.1) lead to further increase of the galactic magnetic field. This further amplification is again much more efficient for the more massive galaxy models of SQ model A. At the time of the present-day configuration, the galactic magnetic field strengths still show an increasing trend and thus have not yet reached the saturation levels (dynamic equilibrium or equipartition between turbulent and magnetic energy density) of the galactic magnetic field within the different models. We note that the total galactic magnetic field is a mean value of all four galaxies which had undergone interactions of different intensity and violence. The IGM magnetic field of both SQ models gets only slightly amplified during the encounters and reaches an value of order $10^{-9}$ to $10^{-8}$G, depending on the simulation. Again, the interactions of the more massive SQ galaxy models A lead to a more efficient amplification of the IGM magnetic field.

The radial profiles (with the origin being the centre of the main galaxy NGC 7319) of the mean, minimum and maximum magnetic field strength within a radius of 260 Mpc are shown in Fig. \ref{magn1} for SQ model A and in Fig. \ref{magn2} for SQ model B, respectively, at different evolutionary stages. Both simulations show an amplification of the magnetic field at small radii, i.e. in the region where the galaxies are located, prior to the first interaction (A0 and B0, respectively). This amplification is due to the winding process. At larger radii, the magnetic field strengths stay constant at the value of the initial IGM magnetic field ($10^{-9}$G). In both models, the magnetic field strength at smaller radii, i.e. within the galaxies, gets efficiently amplified during the interactions. As before, for the more massive SQ model A, the amplification is more efficient. Also, within SQ model A, the higher magnetic field is transported farther out into the IGM due to the stronger outflows and thus a stronger dilatation of the shocked regions.

\subsection[]{Radio emission and polarization maps}

We compute synthetic radio emission and polarization maps for our simulations of the two different representations of SQ in an analogous way as presented in \citet{KoLe11}. The calculation is outlined in Appendix A. 

\begin{figure}
\includegraphics[width=1.\columnwidth]{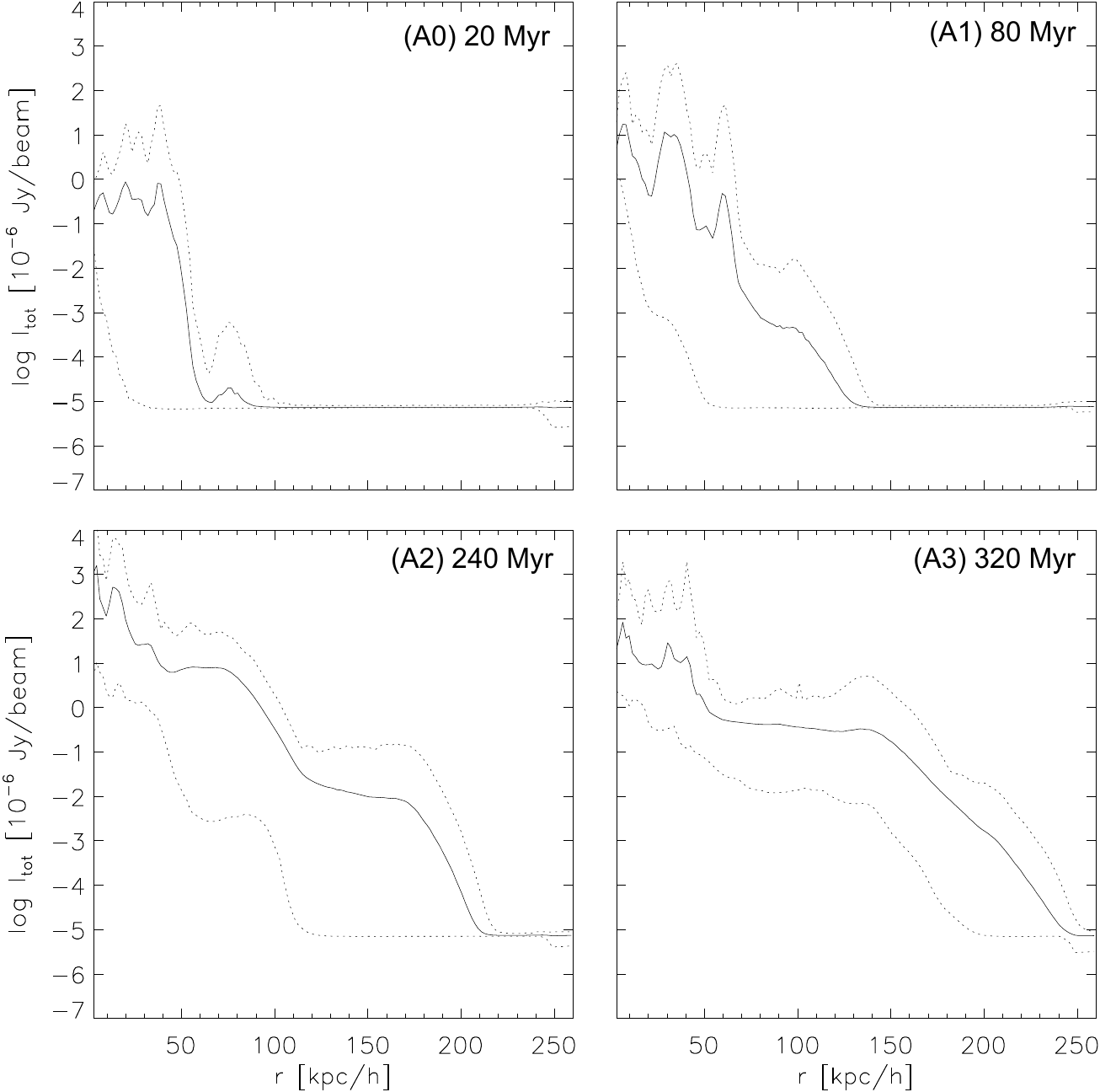}
\caption{\small{Radial profiles (with the origin being the centre of the main galaxy NGC 7319) of mean (solid lines), maximum (upper dotted lines) and minimum (lower dotted lines) synchrotron emission at $\nu=4.86 \cdot 10^{9}$ Hz for SQ model A at different evolutionary stages. }}
\label{sync1}
\vspace{0.01cm}
\includegraphics[width=1.\columnwidth]{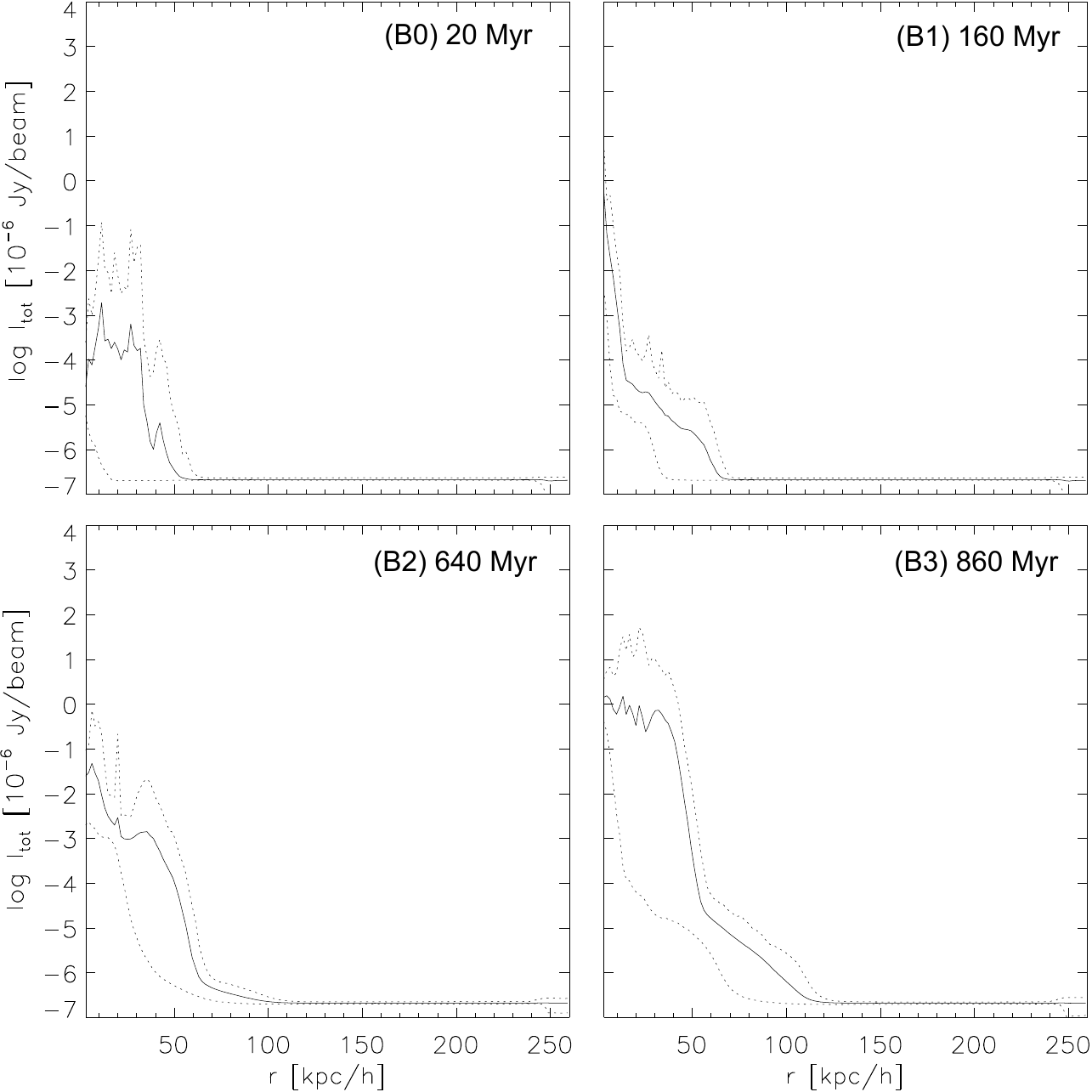}
\caption{\small{Same as Fig. \ref{sync1}, but for SQ model B. Within both models, the total synchrotron intensity gets amplified during the interactions. The amplification is more efficient for the more massive SQ model A.}}
\label{sync2}
\end{figure}

Fig. \ref{radiotot} shows the synthetic radio maps where contours of the total intensity are overlaid on a density color plot (we use the same color code as in Fig. \ref{densAB}a and  Fig. \ref{densAB}b, respectively). The left panels  (a, c, e)  show the present-day configuration of the SQ model A, and the right panels  (b, d, f) the present-day configuration of the SQ model B. The radio emission is calculated for three different frequencies $\nu=4.86 \cdot 10^{9}$ Hz, $\nu=1.4 \cdot 10^{9}$ Hz and $\nu=0.2 \cdot 10^{9}$ Hz (from top to bottom). For the SQ model A, the contour levels of the radio emission are similar to the levels used by \citet{XuLu03} (cf. also Fig. \ref{Xu}). As SQ model B shows a much lower radio emission, the contour levels are different. In all panels, we highlight the 50 $\mu$Jy/beam level with a pink contour.

The synthetic radio maps of SQ model A (left panels in Fig. \ref{radiotot}) reveal a high total synchrotron intensity in the central part of NGC 7319, in the disc of NGC 7318a and in the region between NGC 7319 and the galaxy pair NGC 7318a/b. The region of enhanced synchrotron intensity east of NGC 7318b coincides very well with the shock region found in the observed X-ray luminosity. The regions of high synchrotron emission correspond well to observations \citep[cf. Fig. \ref{Xu}, and][their Fig. 4]{XuLu03}. There is a further region of enhanced synchrotron emission north of NGC 7318b, which is presumably due to a strong outflow. The strength of the synchrotron intensity corresponds to the observed order of magnitude, but however, a direct comparison of the absolute strength of the synchrotron intensity is rather difficult due to the simplifying assumption of the cosmic ray (CR) energy distribution (cf. Appendix A).

The synthetic radio maps of SQ model B (right panels of Fig. \ref{radiotot}) reveal a high total synchrotron intensity within the disc of NGC 7319 and a slightly lower synchrotron emission around NGC 7318a. Also, there is a region of high synchrotron emission between NGC 7319 and NGC 7318a/b, which coincides with the shock region found in the X-ray emission. In general, the regions of enhanced synchrotron emission agree well with observations \citep[cf. Fig. \ref{Xu}, and][their Fig. 4]{XuLu03}. However, the strength of the synchrotron emission within this SQ model is approximately a factor of 100 lower compared to SQ model A. There is a region of slightly enhanced synchrotron intensity in the north of the galaxy pair NGC 7318a/b, which may be compared to the observed star-forming region also showing high radio emission \citep{XuLu03}.

Within both models, the contour level of 50 $\mu$Jy/beam (highlighted in pink) encloses larger areas for lower frequencies. This behaviour is caused by two cumulative effects: First, the assumed power-law spectrum of the CR electrons (see Appendix A) results in fewer CR electrons with high energies compared to lower energies. The CR electrons with high energies are responsible for the high-frequency synchrotron emission, whereas the CR electrons with lower energies lead to a low-frequency emission. Consequently, the low frequency emission should be more intense (more photons) than the high frequency emission. Second, the strongest magnetic fields develop in the region of the large shock, the galaxies and the outflows. Thus, the  high-frequency synchrotron emission should be concentrated in these areas. Together, this leads to an overall low intensity of the high-frequency emission. On the other hand, there are wide areas of weak magnetic field producing a high amount of low-frequency emission. Thus, the low-frequency emission is at last more intense. Applying smaller frequencies, we find a slightly higher visibility of details in the synchrotron distribution. But generally the regions of enhanced synchrotron emission remain comparable to the observations of \citet[][cf. their Fig. 4]{XuLu03} (cf. Fig. \ref{Xu}).

Fig. \ref{radiopol} shows the synthetic radio map of model A (a) and model B (b), respectively, at 4.86 GHz for the present-day configuration (A3 and B3, respectively, cf. section 5.1). Colors visualize the total intensity $I_{tot}$ and the black contours show the polarized Intensity $I_{p}$ (in $\mu$Jy/beam). The direction of the magnetic field is indicated by the white lines, whereby the length of these lines is scaled according to the degree of polarization $\Pi_{obs}$ (cf. length scale in lower left corner of the plots). Total as well as polarized synchrotron emission is visible inside but also outside the galactic discs, indicating that the interaction-driven shocks have already magnetized the sourrounding IGM. In both models, high total and polarized synchrotron emission is found within the region of the large shock between NGC 7319 and NGC 7318a/b. The contour levels of the polarized synchrotron emission within SQ model B are approximately a factor of 100 lower compared to SQ model A.

\begin{table*}
 \begin{minipage}{170mm}
\caption{Achievement of our SQ simulations compared to previous numerical studies of SQ}
\begin{center}
\renewcommand{\arraystretch}{1.2}
\scriptsize{
\begin{tabular}{lllll}
\hline\hline
&SQ model by&SQ model A&SQ model by&SQ model B\\
&\citet{ReAp10}&&\citet{HwSt12}&
\\\hline
SQ large-scale morphology&qualitative agreement&qualitative agreement&qualitative agreement&qualitative agreement\\
&with observations&with observations&with observations&with observations\\\\

Presence of large shock&-&clearly visible in X-ray&indicated by gas distribution&clearly visible in X-ray\\
&&and radio emission&and shock-heated gas&and radio emission\\\\

Star forming regions&-&general agreement&hardly comparable&correspond roughly\\
&&with observations&(very roughly displayed)&with observations\\\\

Star formation rates&-&behaviour in agreement with&-&behaviour in agreement with\\
&&previous numerical studies&&previous numerical studies\\\\

Gas temperature&-&behaviour in agreement with&hardly comparable&behaviour in agreement with\\
&&previous numerical studies&(very roughly displayed)&previous numerical studies\\\\

X-ray emission regions&-&correspond generally&-&correspond generally\\
&&with observations&&well with observations\\\\

X-ray emission strengths&-&too low compared to&-&too low compared to\\
&&observations&&observations\\\\

Magnetic field&-&behaviour in agreement with&-&behaviour in agreement with\\
&&previous numerical studies&&previous numerical studies\\\\

Radio emission regions&-&good agreement with&-&correspond generally to\\
&&observations&&observations\\\\

Radio emission strengths&-&order of magnitude corresponds&-&low compared to\\
&&to observations&&observations\\
\hline\hline
\end{tabular}
}
\end{center}
\end{minipage}
\label{achiev}
\end{table*}

The radial profiles (with the origin being the centre of the main galaxy NGC 7319) of the mean, minimum and maximum total synchrotron emission at $\nu=4.86 \cdot 10^{9}$ Hz within a radius of 260 Mpc are shown in Fig. \ref{sync1} for SQ model A and in Fig. \ref{sync2} for SQ model B, respectively, at different evolutionary stages. The profiles of the synchrotron intensity evolve similar to the profiles of the magnetic field (Figs. 10 and 11): Within both models there is an amplification of the total synchrotron emission at small radii prior to the first interaction (A0 and B0). At larger radii, the total synchrotron emission stays constant, which corresponds to the constant initial IGM magnetic field strength and orientation in the outer regions. During the interactions, the total synchrotron intensity gets amplified within both SQ models. At the time of the second and third interaction (A2 and A3, B2 and B3), the synchrotron emission is already entering larger radii, i.e. the IGM. In case of the more massive SQ model A, the initial synchrotron intensity within the galaxies and the IGM is higher and also the amplification is more efficient compared to SQ model B. Also, the synchrotron emission is transported farther outwards into the IGM.

%%%%%%%%%%%%%%%%%%%%%%%%%%%%%%%%%%%%%%%%%%%%%%%%%%%%%%%%%%%%%%%%%%%%%%%%%%%%%%%%%%%%%%%%%%%%%%%%%%%%%%%%%%%%%%%%%%%%%%%%%%
%%%%%%%%%%%%%%%%%%%%%%%% Discussion of Models %%%%%%%%%%%%%%%%%%%%%%%%%%%%%%%%%%%%%%%%%%%%%%%%%%%%%%%%%%%%%%%%%%%%%%%%%%%%
%%%%%%%%%%%%%%%%%%%%%%%%%%%%%%%%%%%%%%%%%%%%%%%%%%%%%%%%%%%%%%%%%%%%%%%%%%%%%%%%%%%%%%%%%%%%%%%%%%%%%%%%%%%%%%%%%%%%%%%%%%

\section{Discussion of the SQ models}

The comparison of our numerical simulations with observations of SQ yields some agreement with several observed properties of the gaseous and stellar components of SQ. However, the quality of the match with observations is different for our two SQ models. In Table 7 we present a brief listing of the achievement of our simulations in comparison with the results of the previous numerical studies of SQ by \citet{ReAp10} and \citet{HwSt12}.

In general, both presented models of SQ are capable of reproducing its large-scale structure, but the spacial distribution of matter is different due to the different sizes of the participating galaxies. Furthermore, the enhancement of gaseous properties is naturally more efficient for SQ model A compared to SQ model B, which is mainly due to the larger total masses of the involved progenitor galaxies.

Regions of high synchrotron intensity might correspond to regions of high density, however, within periods of intensive shock ejection, the magnetic field is amplified by turbulence and shocks. During the interactions, the magnetic field and correspondingly the total and polarized synchrotron intensities are enhanced behind the shocks and transported into the IGM. This behaviour is well displayed in our SQ simulations.

The observed prominent shock within SQ develops within both of our SQ models. The shock region is found in both the synthetic X-ray and the radio maps of our simulations. The distribution of the total synchrotron intensity compares very favorably with observations \citep[cf. Figs. 2 and 12, see also][]{XuLu03}. The highest synchrotron values are reached within the region of the large shock in both SQ models. However, model A reproduces the strength of the synchrotron emission in the observed order of magnitude, whereas the extension of the shock region is slightly better reproduced by model B, which clearly underestimates the strength of the synchrotron emission. 

The final configuration in SQ model A is reached after the high-speed intruder NGC 7318b hits the system and finally collides with NGC 7318a. Nevertheless, we claim that the large shock within this system results from the collision of NGC 7318b with the IGM. Especially the direction of the magnetic field vectors (which are directing towards NGC 7318b) supports this origin of the shock, which is also consistent with general acceptance. However, it cannot be ruled out that the large shock in SQ model A results at least partially from a collision of NGC 7318b with NGC 7318a. On the contrary, the large shock in SQ model B is certainly a result of an interaction of NGC 7318b with the IGM as the interaction between NGC 7318a and NGC 7318b happened approximately 220 Myr before the present-day configuration is reached.

Overall, we can recognize a general trend in our simulations: the enhancement and propagation of regions with higher values of the studied gaseous properties (i.e. temperature, X-ray emission, magnetic field strength and synchrotron intensity) resulting from interactions and associated shocks and outflows is much more efficient for the more massive galaxies of SQ model A. SQ model B shows qualitatively similar effects of enhancement and propagation, but to a smaller extent.
This is not surprising, as higher masses lead to higher kinetic energies of the interacting galaxies, which results in higher equipartition values i.e. the thermal energy (temperature) or the magnetic energy (magnetic field strength).
These general findings and the good agreement of the synchrotron intensity in model A with observations may be interpreted as an indication for larger progenitor galaxies within SQ. Therefore, the promising extension of the shock region in both X-ray and synchrotron emission motivating the underlying formation scenario of SQ model B in combination with the total masses of the SQ model A would provide a very good starting basis for further studies.

%%%%%%%%%%%%%%%%%%%%%%%%%%%%%%%%%%%%%%%%%%%%%%%%%%%%%%%%%%%%%%%%%%%%%%%%%%%%%%%%%%%%%%%%%%%%%%%%%%%%%%%%%%%%%%%%%%%%%%%%%%
%%%%%%%%%%%%%%%%%%%%%%%% Conclusion and Outlook %%%%%%%%%%%%%%%%%%%%%%%%%%%%%%%%%%%%%%%%%%%%%%%%%%%%%%%%%%%%%%%%%%%%%%%%%%
%%%%%%%%%%%%%%%%%%%%%%%%%%%%%%%%%%%%%%%%%%%%%%%%%%%%%%%%%%%%%%%%%%%%%%%%%%%%%%%%%%%%%%%%%%%%%%%%%%%%%%%%%%%%%%%%%%%%%%%%%%

\section{Conclusions}
We have presented simulations of Stephan's Quintet including magnetic fields, radiative cooling, star formation and supernova feedback. We have investigated different properties of the gaseous component for two different galaxy models based on \citet{ReAp10}, SQ model A, and based on \citet{HwSt12}, SQ model B, respectively. We have set the focus on the general morphology, on the distribution of star-forming regions and star formation rates, on the temperature and the corresponding X-ray emission and finally on magnetic fields and the resulting total and polarized radio emission. A brief listing of the achievement of our simulations in comparison with the previous studies by \citet{ReAp10} and \citet{HwSt12} is shown in Table 7. The main results of our simulations can be summarized as follows:

\begin{itemize}
 \item The present-day configuration of SQ model A develops within 320 Myr. The morphology of the system agrees qualitatively well with observations, only the position of the galaxy pair NGC 7318a/b, its small-scale details and the inner and outer tails cannot be reproduced correctly. The outer tail is generated in this model but already too diffuse to be visible at the present-day configuration as already noted for the original model of \citet{ReAp10}.

 \item The present-day configuration of SQ model B develops within 860 Myr. Again, the morphology of the system agrees qualitatively well with observations, however, the position of NGC 7318a is slightly too southern. Also, the small-scale features such as the arms of NGC 7318b or the smaller-scale structure of NGC 7319 cannot be reproduced correctly and the outer tail is shorter compared to observations.

 \item Within SQ model A, the total masses of the galaxies are approximately Milky Way-like. In contrast, the galactic masses of SQ model B are roughly a factor of 10 smaller compared to SQ model A. As lower galactic masses imply lower equipartition energies, the enhancement of the gaseous properties is commonly lower for SQ model B.

 \item The regions of active star formation within SQ model A are found mainly in the discs of the galaxies, and also within the inner tail and between NGC 7319 and the pair NGC 7318a/b. The latter partly coincides with the region of the large shock. Within SQ model B, the regions of active star formation are found within the inner discs of NGC 7319 and the galaxies NGC 7318a/b, but there is no region of active star formation between these galaxies.

 \item The global SFR strongly depends on the initial masses of the galaxies and is significantly higher for the more massive SQ model A.

 \item In both models, the temperature of the gas within the galaxies is cooler compared to the IGM, which gets heated by shocks and outflows caused by the interactions. The mean temperature in SQ model B is significantly lower compared to SQ model A.

 \item The X-ray emission shows the highest luminosities in the region of the large shock between NGC 7319 and the pair NGC 7318a/b within both models, in good agreement with observations \citep{PiTr97,SuRo01}. The X-ray luminosity in the shock region within SQ model B is about one order of magnitude smaller compared to SQ model A.

 \item We find high values of the magnetic field strength in the region of the large shock and also within outflow regions in both SQ models. The values of the magnetic field strength within SQ model B are approximately a factor of 3 smaller compared to SQ model A.

 \item The temporal evolution of the mean total magnetic field reveals an amplification of the magnetic field strengths along with the different interactions. Thereby, the increase in the magnetic field strength is more efficient for SQ model A.

 \item The synthetic radio maps of both models show a high total and polarized synchrotron intensity within the large shock, within NGC 7319 and around and within NGC 7318a. This finding agrees well with observations \citep[cf.][]{XuLu03}.
\end{itemize}

The large shock revealed by observations of SQ is most likely the result of a collision of NGC 7318b with the IGM. The observed ridge of radio emission can therefore be ascribed to shock activity. The shock front in our simulations is clearly visible in the X-ray and synchrotron emission within both SQ models. We emphasize the importance of shocks for the magnetic field amplification and the enhancement of the synchrotron emission. Whenever a  high amount of synchrotron emission is detected in regions between interacting galaxies, it may be ascribed to shock activity.

For future studies, a further development of the existing SQ models would be essential to draw more detailed conclusions on the extension and strength of the synchrotron emission within SQ. As the SQ model B results in a lower enhancement of the gaseous properties mainly because of the smaller masses, but displays the regions of enhanced X-ray and synchrotron emission quite well, it would be worthwile to use a different scaling of the total masses of SQ model B comparable to the total masses of the SQ model A. This would lead to a better comparability of the strengths of the gaseous properties of the present-day configuration of the two different models of SQ. Another particular focus in further studies should thereby be placed on the position and extension of the galaxy pair NGC 7318a/b, which we found to significantly affect the extension and structure of the large shock in SQ. As in our simulations the used particle masses are of the order of the mass of the largest molecular clouds, small-scale turbulence within the large shock region as recently observed by \citet{Gu12} cannot be modeled in our work. Therefore, further numerical simulations focusing on smaller scales would lead to a deeper understanding of the involved processes of shock activity, especially shocks wrapped around clouds and cloud like structures. Furthermore, observations of the radio emission ridge at different frequencies would be of particular interest in order to gain new insights into the shock region. This knowledge could then be used as a basis for further improvements of numerical SQ models.

%%%%%%%%%%%%%%%%%%%%%%%%%%%%%%%%%%%%%%%%%%%%%%%%%%%%%%%%%%%%%%%%%%%%%%%%%%%%%%%%%%%%%%%%%%%%%%%%%%%%%%%%%%%%%%%%%%%%%%%%%%
%%%%%%%%%%%%%%%%%%%%%%%% Acknowledgements %%%%%%%%%%%%%%%%%%%%%%%%%%%%%%%%%%%%%%%%%%%%%%%%%%%%%%%%%%%%%%%%%%%%%%%%%%%%%%%%%%%%%%%
%%%%%%%%%%%%%%%%%%%%%%%%%%%%%%%%%%%%%%%%%%%%%%%%%%%%%%%%%%%%%%%%%%%%%%%%%%%%%%%%%%%%%%%%%%%%%%%%%%%%%%%%%%%%%%%%%%%%%%%%%%

\section*{Acknowledgments}
We thank the referee Jeong-Sun Hwang for very helpful comments which improved the paper significantly. AG is grateful for interesting and helpful discussions with Harald Lesch, Dominik Bomans and Federico Stasyszyn and thanks Volker Springel for the programs to set up the initial galaxy models.
Rendered plots were made using \textsc{P-Smac2} (Donnert et al., in preparation). Granting of computing time from John von Neumann-Institute for Computing (NIC), J\"{u}lich, Germany, is gratefully acknowledged. We acknowledge support through the DFG Research Unit 1254.
KD acknowledges the support by the DFG Priority Programme 1177 and additional support by the DFG Cluster of Excellence \textquoteleft Origin and Structure of the Universe\textquoteright.

\appendix
\section{Calculation of total and polarized synchrotron emission}

The calculations of the total and polarized synchrotron emission were performed using the code \textsc{P-Smac2} (Donnert et al., in preparation), which projects the corresponding values on a grid. For convenience of the reader, the required calculations are outlined below.

The radiation spectrum can be evaluated for a given distribution of electron energies. The energy spectrum of the cosmic ray electrons can be approximated by a power-law:
\begin{equation}
n(E) \text{d}E = \kappa E^{-p} \text{d}E,
\end{equation}
with $n(E)$ being the number density of electrons and $p=2.6$ the index of the power spectrum. The latter implies a spectral index $\alpha = (p-1)/2$ of 0.8, which was also found by \citep{XuLu03} for NGC 7319. $\kappa$ is a constant normalization factor of the cosmic ray energy spectrum, which can be calculated for a given CR energy density $e_{CR}$:
\begin{equation}
e_{CR}=\int_{E_{min}}^{E_{max}} En(e)\text{d}E = \kappa \int_{E_{min}}^{E_{max}} E^{(1-p)}\text{d}E,
\end{equation}
with an assumed energy range of $E_{min} = 10^{9}$ eV to $E_{max} = 10^{15}$ eV. The CR energy distribution is expected to be proportional to the thermal energy distribution. We assume $e_{CR}=0.01 \cdot e_{therm}$. 

We calculate the total synchrotron emission $J_{\nu}$ at a given frequency $\nu$ via \citep{Lo11}
\begin{eqnarray}
\hspace*{3mm}J_{\nu} = \frac{\sqrt{3} e^{3} B_{\bot} \kappa}{m_{e} c^{2} (p+1)}  \left( \frac{m_{e}^{3} c^{5} 2\pi \nu}{3 e B_{\bot}}\right)^{-\frac{p-1}{2}}\hspace*{-4mm}\nonumber \\
\hspace*{3mm} \times \Gamma \left( \frac{p}{4} + \frac{19}{12}\right) \Gamma \left( \frac{p}{4} - \frac{1}{12}\right) \hspace{2mm} \left[ \text{Jy}\right] ,
\end{eqnarray}

with the Gamma function $\Gamma$, the electron mass $m_{e}$, the electron charge $e$ and the speed of light $c$. The frequency is set to $\nu=4.86 \cdot 10^{9}$ Hz and $\nu=1.4 \cdot 10^{9}$ Hz, respectively, which correspond to the frequencies used for the observations presented by \citet{XuLu03} (see Fig. \ref{Xu}). Additionally, we also use a frequency of $\nu=0.2 \cdot 10^{9}$ Hz. $B_{\perp}$ denotes the magnetic field component perpendicular to the line of sight and is taken from the simulations. 

Within each grid cell, the total synchrotron emission $J_{\nu}$ is calculated. Integration of $J_{\nu}$ along the line-of-sight gives the total synchrotron intensity $I_{tot}$. The polarized emission $I_{pol}$ can be obtained using the Stokes parameters $Q$ and $U$:
\begin{eqnarray}
I_{pol}=\sqrt{Q^{2}+U^{2}} \hspace*{2cm}\nonumber \\
= \sqrt{\left( \Pi \int_{los} J_{\nu} \cos(2\psi) \text{d}s\right) ^{2} + \left( \Pi \int_{los} J_{\nu} \sin(2\psi) \text{d}s\right)^{2} },
\end{eqnarray}
where the integration is carried out over the line-of-sight (los). In case of a power-law form of the energy spectrum, the degree of polarization can be evaluated using \citep{Lo11}
\begin{equation}
\Pi = \frac{p+1}{p+\frac{7}{3}}.
\end{equation}
The polarization angle $\psi$ is defined as the angle between the electric field vector $\overrightarrow{E}_{\perp}$ of the radiation perpendicular to the magnetic field and the $x$-axis in the plane of the sky ($xy$-plane). Thus, $\sin(2\psi)$ and  $\cos(2\psi)$ can be calculated as follows:
\begin{equation}
\sin(2\psi)=-\frac{2 B_{x}B_{y}}{B_{x}^{2}+B_{y}^{2}}, \hspace{3mm} \cos(2\psi)=-\frac{B_{x}^{2}-B_{y}^{2}}{B_{x}^{2}+B_{y}^{2}}.
\end{equation}
The observed degree of polarization $\Pi_{obs}$ can finally be calculated using:
\begin{equation}
\Pi_{obs}=\frac{I_{pol}}{I_{tot}}.
\end{equation}
To avoid errors due to the spatial isotropy of the emission, the total and polarized intensities are multiplied with a factor
\begin{equation}
f_{obs}=\frac{\pi \cdot r^{2}_{\text{beam}}}{4\pi \cdot d^{2}},
\end{equation}
with $d$ being the distance to the observer and $r_{\text{beam}}$ the assumed radius of the beam corresponding to a resolution of $20''$. For the distance to SQ, we use the estimated value of 94 Mpc \citep{MoMa98,ApXu06}. The artificial flux then corresponds to the expected amount reaching Earth from the distance of SQ.

\end{document}